\documentclass[a4paper,11pt]{article}
\pdfoutput=1 

\usepackage{jcappub, subcaption} 

\usepackage[T1]{fontenc} 

\title{Neutron stars in $f(\mathcal{R,T})$ gravity using realistic equations of state in the light of massive pulsars and GW170817}

\author[a,b,c]{R. Lobato,}
\author[b]{O. Louren\c co,}
\author[b,d]{P. H. R. S. Moraes,}
\author[b]{C. H. Lenzi,}
\author[b,e]{M. de Avellar,}
\author[b]{W. de Paula,}
\author[b]{M. Dutra,}
\author[b,c]{and M. Malheiro}

\affiliation[a]{Department of Physics and Astronomy, Texas A\&M University-Commerce, TX 75429, USA}
\affiliation[b]{Departamento de F\'\i sica, Instituto Tecnol\'ogico de Aeron\'autica, 12228--900, SJC, SP, Brazil.}
\affiliation[c]{ICRANet, P.zza della Repubblica 10, I--65122 Pescara, Italy.}
\affiliation[d]{Instituto de Astronomia, Geof\'\i sica e Ci\^encias Atmosf\'ericas, USP, 05508--090, SP, Brazil}
\affiliation[e]{Departamento de F\'\i sica, Universidade Federal de S\~ao Paulo, Rua S\~ao Nicolau 210, 09913--030, Diadema/SP, Brazil.}

\emailAdd{ronaldo.lobato@tamuc.edu}
\emailAdd{odilon.ita@gmail.com}
\emailAdd{moraes.phrs@gmail.com}
\emailAdd{chlenzi@ita.br}
\emailAdd{mgb.avellar@gmail.com}
\emailAdd{wayne@ita.br}
\emailAdd{marianad@ita.br}
\emailAdd{malheiro@ita.br}

\abstract{In this work we investigate neutron stars (NS) in $f(\mathcal{R,T})$ gravity for the case
  $R+2\lambda\mathcal{T}$, $\mathcal{R}$ is the Ricci scalar and $\mathcal{T}$ the trace of the
  energy-momentum tensor. The hydrostatic equilibrium equations are solved considering realistic equations of state (EsoS). The NS masses and radii obtained are subject to a joint constrain from massive pulsars and the event GW170817. The parameter $\lambda$ needs to be negative as in previous NS studies, however we found a minimum value for it. The value should be $|\lambda|\lesssim0.02$ and the
  reason for so small value in comparison with previous ones obtained with simpler EsoS is due to
  the existence of the NS crust. The pressure in theory of gravity depends on the inverse
  of the sound velocity $v_s$. Since, $v_s$ is low in the crust, $|\lambda|$ need to be very
  small. We found that the increment in the star mass is less than $1\%$, much smaller than previous ones
  obtained not considering the realistic stellar structure, and the star radius cannot become larger, its changes compared
  to GR is less than $3.6\%$ in all cases. The finding that using several
  relativistic and non-relativistic models the variation on the NS mass and radius are almost the
  same for all the EsoS, manifests that our results are insensitive to the high density part of the
  EsoS. It confirms that stellar mass and radii changes depend only on crust, where the EoS
  is essentially the same for all the models. The NS crust effect implying very small values of
  $|\lambda|$ does not depend on the theory's function chosen, since for any other one the hydrostatic
  equilibrium equation would always have the dependence $1/v_s$. Finally, we highlight that our results indicate that conclusions obtained from NS studies done in modified theories of gravity without using realistic EsoS that describe correctly the NS interior can be unreliable.}
\arxivnumber{}

\begin{document}

\maketitle
\flushbottom

\section{Introduction}\label{sec:int}
We have been collecting several possible pieces of evidence that show that General Theory of Relativity (GR)
breaks down under some specific regimes. There are also intriguing observational
features at galactic and cosmological scales, namely the universe's dark sector. For
example, it is necessary to assume that spiral galaxies are generally filled by invisible or dark
matter to account for their rotation curves' flatness~\cite{rubin/1980, rubin/1985}. The structure
formation at cosmological scales demands, in the hierarchical scenario, for enormous dark matter
haloes~\cite{liddle/1993, abadi/2003, saha/2013, delpopolo/2007}. Not enough, in length scales
larger than clusters of galaxies, the Universe dynamics is dominated by a negative pressure fluid,
namely dark energy, which makes the expansion of the universe to accelerate~\cite{riess/2004,
  abbott/2016}.

At the astrophysical level scales, some observational issues are also persistent. Massive
pulsars~\cite{antoniadis/2013, demorest/2010} have been observed and can hardly be explained within
the GR approach unless one strongly modifies the stellar structure~\cite{grigorian/2016, lenzi/2012,
  li/2012}.

A possible form to account for those observational issues is through extended (modified) gravity theories
(EGTs)~\cite{nojiri/2011, nojiri/2017, nunes/2020}. The simplest way to extend GR is through $f(\mathcal{R})$ gravity~\cite{sotiriou/2010,
  defelice/2010}, for which $f(\mathcal{R})$ is a general function of the Ricci scalar
$\mathcal{R}$. Within the metric formalism, such a theory has already shown to be capable of accounting for the acceleration
of the universe expansion with no need for dark energy~\cite{cognola/2008, navarro/2007,
  song/2007}. However, the solar system regime seems to rule out most of the $f(\mathcal{R})$ models
proposed so far~\cite{chiba/2003, erickcek/2006, capozziello/2007a, capozziello/2008a,
  olmo/2007}. This theory also was applied to neutron stars (NS)~\cite{astashenok/2013,
  astashenok/2015a, capozziello/2016, astashenok/2020}. Nevertheless, the existence of $f(\mathcal{R})$ gravity singularities could forbid the
formation of such objects \cite{kobayashi/2008}.

Anyhow it is worth to remark that $f(R)$ gravity was also developed in the Palatini
  formalism \cite{sotiriou/2006, olmo/2011} and within this context the theory may
  present optimistic results in the solar system regime, as it was recently shown in
  \cite{toniato/2020}, and also in what concerns compact stars stability \cite{bhatti/2019, kainulainen/2007}.

In the present paper, we shall investigate the hydrostatic equilibrium configurations of stellar
objects within EGTs that allow the \textit{rhs} of Einstein's field equations to be generalized rather
than their \textit{lhs}, as in $f(\mathcal{R})$ gravity.

We will assume the $f(\mathcal{R,T})$ theory as our underlying theory of gravity in the present
work. The $\mathcal{T}$ stands for the trace of energy-momentum tensor and the
$\mathcal{T}$-dependence motivation in such a scenario is related to the possibility of imperfect
fluids to exist in the universe or to quantum effects~\cite{lobato/2019}.

It should also be mentioned that here we work with the metric formalism of the $f(R,T)$
  gravity, but this theory has also been developed in the Palatini formalism \cite{wu/2018} and some of its applications can be seen in~\cite{barrientos/2016,bhatti/2019b}.

There are some interesting outcomes obtained from $f(\mathcal{R,T})$ gravity
applications. For instance, we quote Reference~\cite{carvalho/2017}, in which some of us showed
that the $f(\mathcal{R,T})$ gravity might increase the maximum masses of white dwarfs, getting in
touch with some observational data.

Notably, we are also going to investigate the stellar equilibrium configurations in the present
paper, but rather, the equilibrium configurations of neutron stars (NSs). NSs are supernova remnants
known for their high density, strong gravitational field, and rapid rotation
rate~\cite{lattimer/2001,lattimer/2004}. Their relevance has recently increased in both theoretical
and observational aspects.  Besides the aforementioned massive pulsars, NSs have been vital sources
of detected gravitational waves~\cite{abbott/2017, abbott/2020}.

The understanding of stellar structure from the modified gravity perspective and the properties of
strongly interacting matter at ultra-high densities provide new phenomenological predictions. With
recent observations we have a window to constrain parameters coming from both sides.

The hydrostatic equilibrium configurations of NSs in the $f(\mathcal{R,T})$ gravity has been
investigated in Ref.~\cite{moraes/2016} from a simple barotropic equation of state (EoS) describing
matter inside these objects. Here, we intend to be more
rigorous than~\cite{moraes/2016} and will consider the stellar structure of neutron stars
  and apply for the first time in this theory a set of fundamental nuclear matter equations of state
  based on effective models of nuclear interactions, considering non-relativistic and relativistic
  cases, and by comparing our results with gravitational-wave observations, particularly concerning
  to GW170817 event~\cite{abbott/2017}, and also with massive pulsars in a joint constrain.

Consider the stellar structure, we mean, take in account that neutron stars contain matter at
densities from few $\mathrm{g/cm^3}$ at their surface to more than $10^{15}~\mathrm{g/cm^3}$ at the
center; due to this change in the stellar density, the composition changes as one moves from the
center to the surface, i.e., the EoS changes. According to the current theories, a NS can be
subdivided into the atmosphere (where we have a plasma region governed by very intense
magnetic/electric fields) and additional four regions: the outer crust, the inner crust, the outer
core, and the inner core. So, to describe all these different stellar layers, we need different
theories: plasma physics, atomic structure, and nuclear many-body theories in the high
density-temperature regime for the outer region (outer and inner crust); for the inner and outer
cores, we need many-body theories of high dense strongly interacting systems, for details see \S1.3
of the book \textit{Neutron Stars 1} by Haensel \textit{et al.}~\cite{haensel/2007}.

Due to all these different regimes/densities, only the outer crust can be described with accuracy
(this description can the compared with experimental data of atomic nuclei). The EoS describing the
NS interior above nuclear matter density is not yet constrained, being an open question in
astrophysics. However, there are some constraints from microscopic physics such as electric
neutrality, beta equilibrium, and others to describe the interior of neutron star realistically:
causality (the speed of sound, $v_s$, must be less than the speed of light, $c$) and the Le
Chateliers' principles $p\geq0$ and $dp/d\rho>0$.

This uncertainty in the description of the NS interior leads to a large variety of EsoS in the literature, and
they can be separated in \textit{soft} and \textit{stiff} concerning the compressibility of the
nuclear matter and their behavior at high densities, i.e., how fast the pressure changes when the
energy density changes. They also can be divided by the matter compostion: for the outer core, a
$npe\mu$ (neutron-proton-electron-muon) plasma; for the inner core, several possibilities exist such
fermion/boson condensates, hyperons, pion/kaon condensation, or a strange quark star matter at the
star core (around 3 km depending on the quark matter model) surrounded by hadronic matter, this last
possibility is named as hybrid neutron stars in the literature.

The several methods to calculate the EsoS are based on perturbation expansion within the
Brueckner-Bethe-Goldstone theory, perturbation expansion within Green's-function theory, variational
method, effective energy-density functionals, and relativistic mean-field (RMF) models~\cite{ring/1980,
  machleidt/1989, akmal/1997, bethe/1971, haensel/2007}.

Point-coupling (or zero range)
  models~\cite{nikolaus/1992, friar/1996} are also used to describe finite nuclei and nuclear
  matter, as well as nonrelativistic models such as Skyrme~\cite{skyrme/2006, bell/1956,
    skyrme/1958} and Gogny ones~\cite{decharge/1980, berger/1991}. The former is a model in which
  the nucleon-nucleon potential can be written as a contact interaction, and the latter consists of
  a density-dependent zero range term along with two finite range ones (Gaussian type) generating a
  particular momentum dependence in the interaction. By computing only Skyrme and RMF models, it is
  possible to find more than 500 parametrizations. This large number of possibilities naturally
  raises the doubt whether all of them can reproduce different nuclear environments
  simultaneously. In order to start to answer this question, it was studied by some present authors
  in Ref.~\cite{dutra/2012} the capability of 240 Skyrme parametrizations in describing different
  criteria related to the nuclear matter in the vicinity of nuclear saturation density. It was found
  that only 16 satisfy all the constraints simultaneously. A complement of this study was performed
  also in Ref.~\cite{dutra/2014} by some of us, in which 263 parametrizations of the RMF model that
  we are going to use and other ones were tested against an updated set of constraints related to
  nuclear matter, pure neutron matter (PNM), symmetry energy and its derivatives. They include
  limits in the density dependence of the pressure in the symmetric nuclear matter (SNM), coming
  from the experimental data on the motion of ejected matter in the energetic nucleus-nucleus
  collisions; limits on the incompressibility at saturation density in SNM;~limits in the
  low-density region of the energy per particle in PNM;~limits on the symmetry energy at saturation
  density, obtained from isospin diffusion, neutron skins, pygmy dipole resonances, and other
  investigations; among other ones. The detailed description of these constraints is found in
  Ref.~\cite{dutra/2014}.

The outcome of the analysis performed in Ref.~\cite{dutra/2014} is that among the 263
parametrizations of the RMF models studied, only 35 satisfy the updated constraints simultaneously:
BKA20;22;24~\cite{agrawal/2010}, IU-FSU~\cite{fattoyev/2010}, BSR8--12;15--20~\cite{dhiman/2007}, FSU-III-IV~\cite{cai/2012},
FSUGold~\cite{todd-rutel/2005}, G2$^*$~\cite{sulaksono/2006}, FSUGold4~\cite{piekarewicz/2006},
Z271s2--s6~\cite{horowitz/2002}, Z271v4--v6~\cite{horowitz/2002},
FSUGZ03;06~\cite{kumar/2006}. These parametrizations also have been studied in the stellar matter
regime with and without hyperonic matter included, in the context of general
relativity~\cite{dutra/2016}. Some of them can reproduce neutron stars masses around two solar
masses. In Ref.~\cite{lourenco/2019}, the stellar matter was further investigated in GR theory, and
these 35 parametrizations were used to compute the dimensionless tidal deformability ($\Lambda$). In
particular, the interest was to analyze the quantities related to the GW170817 event, in which the
LIGO-VIRGO Collaboration established constraints on $\Lambda$, both for thein a joint constrain from
massive pulsars and the gravitational wave event GW170817 $\Lambda_1,\Lambda_2$ of the two companion
  stars, and for the $\Lambda_{1.4}$ (deformability of the canonical neutron star). Most of the
  consistent RMF parametrizations also satisfy these limits.

The main motivation of this work is to investigate for the first time neutron stars in the
  $f(\mathcal{R,T})$ theory of gravity with realistic hadronic EsoS and considering realistic stellar
  models, that we referred before, and also the case of hybrid neutron stars. We would like to stress that we will use the state of the art of hadronic EsoS, considering a large set of them that have already been restrained. Furthermore, the neutron star masses and radii obtained with these EsoS are subject to a joint constrain from massive pulsars and the gravitational wave event GW170817. We will look for modifications in the neutron star structure (mass and star radius) in this modified theory of gravity and also compare with previous results obtained with an analytical polytropic EoS~\cite{moraes/2016} by some of us. In order to be rigorous in this investigation, we will consider these different generation methods and potentials for the equations of state and exclude those that no longer satisfy the constraints from massive pulsars and the LIGO-VIRGO binary neutron stars observation. For the $npe\mu$ nuclear matter we will consider: (i) the SLy, which is an EoS that uses energy density functional; (ii) the APR1--4, FPS, and WFF1--3, obtained from variational-method;
(iii) the BBB2, which is a nonrelativistic EoS;~ENG and MPA1, which are relativistic, obtained from Brueckner-Hartree-Fock theory; (iv) the BKA20, BSR8, IU-FSU, and Z271s4 which are relativistic mean-field theory EsoS. For the EoS that considers the hybrid matter, we will consider only two parametrizations of ALF, which is a combination of nuclear matter (the crust) and quark matter (the core).

In the next section, we will briefly present the resulting hydrostatic equilibrium equation for
the underlying $f(\mathcal{R,T})$ gravity theory. In Section~\ref{sec:peos}, we will present the piecewise EsoS in the
view of the massive NS observed and in Section~\ref{sec:rmf}, we will present the set of
parametrizations used to describe nuclear and stellar matter constructed from the RMF models. Our results are displayed in Section~\ref{sec:res}, where we investigate in detail the neutron star crust effect, followed by a
careful and in-depth discussion and conclusion of them in Section~\ref{sec:dis}.

\section{Hydrostatic equilibrium equation in $f(\mathcal{R,T})$ gravity}\label{sec:tov}

To work with an EGT that allows the material sector of Einstein's field equations to be generalized
means to have as the starting point an action like~\cite{harko/2011}

\begin{equation}\label{tov1}
S=\int d^{4}x\sqrt{-g}\left[\frac{\mathcal{R}+f(\mathcal{T})}{16\pi}+\mathcal{L}\right],
\end{equation}
where $g$ is the metric determinant, and $f(\mathcal{T})$ is a general function of the trace of the
energy-momentum tensor. Throughout this paper, we assume natural units.

Let us take in Equation~\eqref{tov1}, as the simplest case, $f(\mathcal{T})=2\lambda\mathcal{T}$,
where $\lambda$ is a constant, as done by several authors~\cite{hansraj/2018, deb/2018,
  moraes/2017a, moraes/2017, das/2017}, among many others. In this case, the hydrostatic
equilibrium equation reads~\cite{carvalho/2017, moraes/2016}

\begin{equation}\label{tov2}
  p'=-(\rho+p)\frac{4\pi pr+\frac{m}{r^{2}} -
    \frac{\lambda(\rho-3p)r}{2}}{\left(1-\frac{2m}{r}\right) \left[1+\frac{\lambda}{8\pi+2\lambda}\left(1-\frac{d\rho}{dp}\right)\right]},
\end{equation}
where a prime indicates radial derivative, $m$ is the model-dependent gravitational mass enclosed
within a surface of radius $r$, i.e.,
\begin{equation}
  m' = 4\pi\rho r^{2}+\frac{\lambda}{2}(3\rho-p)r^{2}.
\end{equation}
Moreover, when working with the present formalism, we assumed
$\mathcal{L}=-p$ in~\eqref{tov1}. It is trivial to check that $\lambda=0$ retrieves the standard hydrostatic
equilibrium equation in GR, the so-called TOV (for Tolman-Oppenheimer-Volkoff) equation~\cite{tolman/1939,oppenheimer/1939}.

To solve the system of equations, we will employ EsoS from the piecewise-polytrope representation
used in~\cite{read/2009, raaijmakers/2018}, and also obtained from relativistic mean-field
models~\cite{dutra/2014, dutra/2016}, focusing on the ultra-dense nuclear matter and in the constraints
given by the Laser Interferometer Gravitational-Wave Observatory (LIGO) detections~\cite{theligoscientificcollaborationandthevirgocollaboration/2018, ligoscientificcollaborationandvirgocollaboration/2019}.

\subsection*{Boundary conditions}
The boundary conditions for $f(\mathcal{R,T})$ are the same as in GR, i.e., we have $m(0) = 0,~ p(0) = p_c$
and $\rho(0) = \rho_{c}$ at the center ($r=0$), for which $p_c$ and $\rho_c$ are the central values of the pressure and energy density inside the star, respectively. The stellar surface is the point at radial coordinate $r=R$, where the pressure vanishes, $p(R) = 0$.

\section{Equations of state in view of the massive neutron stars observed}\label{sec:peos}

The piecewise-polytrope representation~\cite{read/2009,
  raaijmakers/2018, carney/2018}, with few parameters, yields macroscopic observables for a
wide range of EsoS. The stellar structure equations map the EoS parameters into the gravitational mass, radius, and moment of inertia. Piecewise EsoS have been extensively used in the context of NSs, and gravitational wave simulations~\cite{lackey/2015, carney/2018, ma/2018, east/2019} and their representation can be tested by
astronomical data, e.g., X-ray data and gravitational waveform.

In our analyses, we used the EsoS from the piecewise-polytrope representation that
yields a maximum mass near $2M_{\odot}$ considering general relativity. Our primary motivation was the two massive observed NS pulsars, namely PSR J0348+0432~\cite{antoniadis/2013} and
PSR J1614--2230~\cite{demorest/2010}, both with $\sim 2M_\odot$. As the upper limit for the mass, we will consider the extremely massive millisecond pulsar recently discovered by Cromartie et
al.~\cite{cromartie/2020}, namely J0740+6620, with $2.14_{-0.18}^{+0.20}~M_{\odot}$ (within 95.4\% credibility
interval). The second criterium is that neutron stars masses and radii obtained by such EsoS are within the mass-radius cloud region delimited by the LIGO-VIRGO observation~\cite{theligoscientificcollaborationandthevirgocollaboration/2018, ligoscientificcollaborationandvirgocollaboration/2019}. Therefore, following these criteria, we obtain NSs which description is consistent with recent astronomical observations.

Let us also remark that the system PSRJ2215+5135, a millisecond pulsar with a mass
$\sim2.27~M_{\odot}$, was also recently observed~\cite{linares/2018}, though the technique used to measure this source is not as precise as those in reference~\cite{cromartie/2020}, (the associated errors are enormous). If these measurements eventually are confirmed with a more precise technique, this pulsar would be one of the most massive neutron stars ever detected.

Moreover, an important observation, just released by the LIGO-VIRGO collaboration, reported a coalescence involving a 22.2--24.3~$M_{\odot}$
black hole and a compact object with 2.50--2.67~$M_{\odot}$, with 90$\%$ credibility~\cite{abbott/2020}. If
this black hole companion is an NS, this could be a breakthrough, since until now no EoS with ordinary matter (i.e., neutrons, protons, electrons) could explain such a
mass in GR context. In this regard, one can check the Figure~\ref{fig:mr_all} below.

Tentatively, there have been proposed some different models of dense matter for stellar objects over the last decades,
such as hyperon, pions-kaons condensation, quarks-strange stars, boson stars, among others. These stars, that could be formed by condensations, strange quarks and bosons stars are still in the theoretical field, and we do not consider them in the Figure~\ref{fig:mr_all}.

\begin{figure}[th]
    \centering
    \includegraphics[scale=0.8]{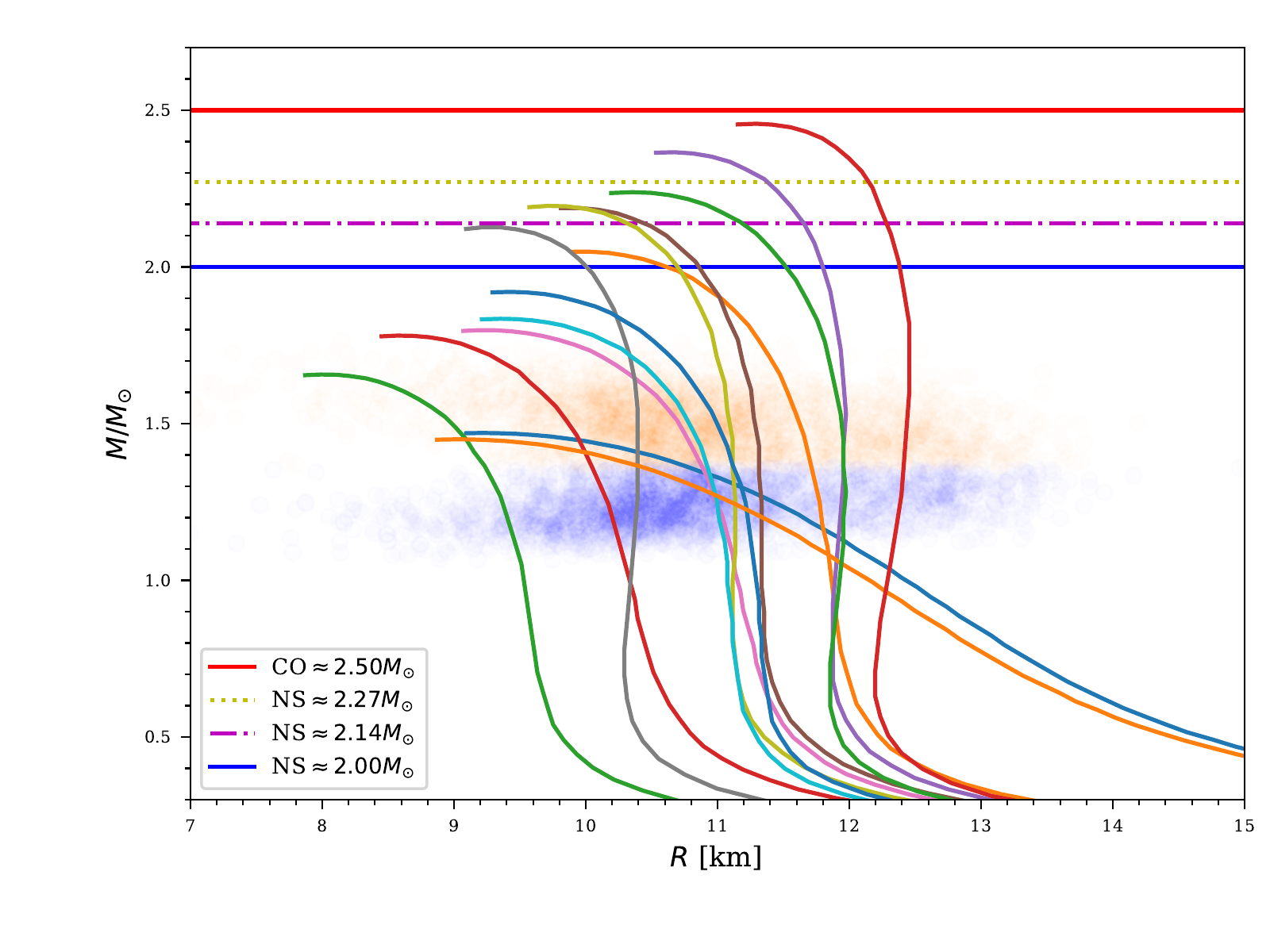}
    \caption{Mass-radius relation in GR for several EsoS considering nucleons. The
      blue and orange regions are the constraints for mass-radius form the GW170817 event~\cite{theligoscientificcollaborationandthevirgocollaboration/2018}. In
      continuous red line we have the minimum mass of the compact object detected by the event GW190414 from the LIGO-VIRGO collaborations~\cite{abbott/2020}.}\label{fig:mr_all}
\end{figure}

To compare the effects on the hydrostatic equilibrium equations, we choose a set of EsoS considering the pure nuclear matter and one EoS for hybrid matter, i.e., with deconfined quarks. They are labeled
according to their name in the literature. For pure nuclear matter, we have non-relativistic
equations of state: APR~\cite{akmal/1998}, which
considers variational-method (VM) with modern nuclear potentials such as Argone and Urbana potentials;
BBB~\cite{baldo/1997}, which is obtained in the framework of the Brueckner-Hartree-Fock (BHF) approximation of the
Brueckner-Bethe-Goldstone (BBG) theory, with realistic two-three particle potentials; the FPS EoS~\cite{lorenz/1993}, being a modern version of the EoS by Friedman
and Pandharipande~\cite{friedman/1981} (FP) it is an EoS which uses the Skyrme model with an energy
density functional (EDF) that considers a nucleon-nucleon interaction by
the Urbana potential and phenomenological three-nucleons interaction; the Skyrme type
SLy EoS~\cite{douchin/2001}, which uses a phenomenological EDF with effective Lyon nuclear interaction of two
potentials, it is similar to the APR one; the WFF EoS~\cite{wiringa/1988} derived from the
variational many-body theory with two-body Urbana potential and a three-body phenomenological
potential (this EoS is also an improvement on the FP one). Concerning relativistic EsoS we consider: ENG~\cite{engvik/1994}, a relativistic Dirac-Brueckner-Hartree-Fock (DBHF) approach, with
modern meson-exchange potential models, and the MPA~\cite{muther/1987}, an extended
relativistic BHF approach for nuclear matter with the exchange of $\pi$ and
$\rho$-mesons. These last two EsoS consider the dependence upon neutron-proton asymmetry.
For EoS containing a hybrid matter of nucleons and quarks, we consider the ALF EoS~\cite{alford/2005}. In this case, the EoS is the combination of nuclear matter (the crust) and quark matter (the core). The crust is described by the APR EoS and the core by a phenomenological parametrization of neutral quarks and an MIT bag model.

From these EsoS, we give special attention to a set of parametrizations: WFF1, APR4, SLy, and MPA1, which are constrained by
the gravitational wave event GW170817~\cite{theligoscientificcollaborationandthevirgocollaboration/2018, ligoscientificcollaborationandvirgocollaboration/2019}. In Most et al.~\cite{most/2018}, further constraints were
obtained using the GW170817 event. For a pure hadronic NS with a mass of 1.4~$M_{\odot}$, the
radii were constrained to be 12.00 < $R_{1.4}/$km < 13.45 with $2\sigma$ confidence, most likely
$\overline{R}_{1.4} = 12.39$~km. Other works of different groups used such an event
to constrain other EsoS as well~\cite{annala/2018}. For details, see Figure 51 in Ref.~\cite{li/2019}. All the EsoS described above are obtained from meson-exchange nuclear potential, and not phenomenological parametrized relativistic mean field hadronic models that we will present in the next section.

Furthermore, constraints from the electromagnetic counterparts of gravitational wave events are
now becoming available~\cite{coughlin/2018, radice/2018a}. These studies have the potential to
constrain the NS matter tightly.

\section{Relativistic mean-field models}\label{sec:rmf}

We also study neutron stars in $f(\mathcal{R,T})$ gravity through a widely known class of parametrizations used
to describe nuclear and stellar matter, namely, those constructed from the so-called relativistic
mean-field models that we already discussed in the introduction. Connecting with the last section concerning the general description of many-nucleon systems (MNS), there are at
least two well-established treatments. One of them is based on a microscopic approach in which a
suitable parametrization of the two-nucleon potential is essential to ensure the reproduction of
some observable, for instance, those related to the deuteron such as its binding energy and
scattering data~\cite{ring/1980, machleidt/1989, akmal/1997}. A way of constructing MNS from the knowledge of the nucleon-nucleon
interaction is from using some methods, such as the Brueckner-Hartree-Fock one~\cite{ring/1980,
  machleidt/1989, akmal/1997, bethe/1971} as we pointed out in the last section. An alternative to these microscopic calculations is the use of phenomenological hadronic models
based on the mean-field approximation. From this specific point of view, the thermodynamic equations
of state of the models are obtained (pressure, energy density, chemical potentials, and others) and the
free constants of each used parametrization are fitted in order to reproduce data from MNS such as
those from finite nuclei or infinite nuclear matter (isotropic system with an infinite number of nucleons
with no spatial boundaries and no Coulomb interaction). Here, we focus on the latter
description and analyze some parametrizations of the finite range relativistic mean-field (RMF)
model given by the following Lagrangian density~\cite{dutra/2014, li/2008},
\begin{align}
\mathcal{L} &= \overline{\psi}(i\gamma^\mu\partial_\mu - M)\psi
+ g_\sigma\sigma\overline{\psi}\psi
- g_\omega\overline{\psi}\gamma^\mu\omega_\mu\psi
- \frac{g_\rho}{2}\overline{\psi}\gamma^\mu\vec{\rho}_\mu\vec{\tau}\psi
+ \frac{1}{2}(\partial^\mu \sigma \partial_\mu \sigma
- m^2_\sigma\sigma^2)
\nonumber\\
&- \frac{A}{3}\sigma^3 - \frac{B}{4}\sigma^4 -\frac{1}{4}F^{\mu\nu}F_{\mu\nu}
+ \frac{1}{2}m^2_\omega\omega_\mu\omega^\mu
+ \frac{C}{4}(g_\omega^2\omega_\mu\omega^\mu){^2}
- \frac{1}{4}\vec{B}^{\mu\nu}\vec{B}_{\mu\nu}
+ \frac{1}{2}m^2_\rho\vec{\rho}_\mu\vec{\rho}^\mu
\nonumber\\
&+ g_\sigma g_\omega^2\sigma\omega_\mu\omega^\mu
\left(\alpha_1+\frac{1}{2}{\alpha'_1}g_\sigma\sigma\right)
+ g_\sigma g_\rho^2\sigma\vec{\rho}_\mu\vec{\rho}^\mu
\left(\alpha_2+\frac{1}{2}{\alpha'_2}g_\sigma\sigma\right)
+ \frac{1}{2}{\alpha'_3}g_\omega^2 g_\rho^2\omega_\mu\omega^\mu
\vec{\rho}_\mu\vec{\rho}^\mu,
\label{rmf}
\end{align}
in which the Dirac spinor $\psi$ is the nucleon field, where $\sigma$, $\omega^\mu$, and $\vec{\rho}_\mu$ represent the scalar, vector and isovector fields related to the mesons
$\sigma$, $\omega$, and $\rho$, respectively. $F_{\mu\nu}=\partial_\nu\omega_\mu-\partial_\mu\omega_\nu$ and
$\vec{B}_{\mu\nu}=\partial_\nu\vec{\rho}_\mu-\partial_\mu\vec{\rho}_\nu$ give the antisymmetric tensors $F_{\mu\nu}$ and
$\vec{B}_{\mu\nu}$. The nucleon mass is $M$,
and the mesons masses are $m_\sigma$, $m_\omega$, and $m_\rho$ ($g_\sigma$, $g_\omega$,
$g_\rho$, $A$, $B$, $C$, $\alpha_1$, $\alpha_1'$, $\alpha_2$, $\alpha_2'$ and $\alpha_3'$ are the
coupling constants). Through the use of the Euler-Lagrange equations, it is possible to obtain the
filed equations of the model. Furthermore, the implementation of the mean-field
approximation~\cite{walecka/1986, walecka/1974} in these equations allows the determination of the energy-momentum
tensor. This quantity determines the energy density ($\rho$) and pressure ($p$) of the model.
All the remaining thermodynamics can be found from $p$ and $\rho$. For detailed calculations, we
address the reader to references~\cite{dutra/2014, li/2008}.

 Here we choose BKA20, BSR8, IU-FSU, and Z271s4 as
representative parametrizations of the ``families'' BKA, BSR, FSU, and Z271, in order
to investigate their predictions on the mass-radius diagram when submitted to the $f(\mathcal{R,T})$
gravity theory. We present our findings in Sec.~\ref{sec:res}.

\section{Results}\label{sec:res}

\begin{figure}[ht]
  \centering
  \begin{subfigure}[b]{0.495\textwidth}
    \includegraphics[scale=0.47]{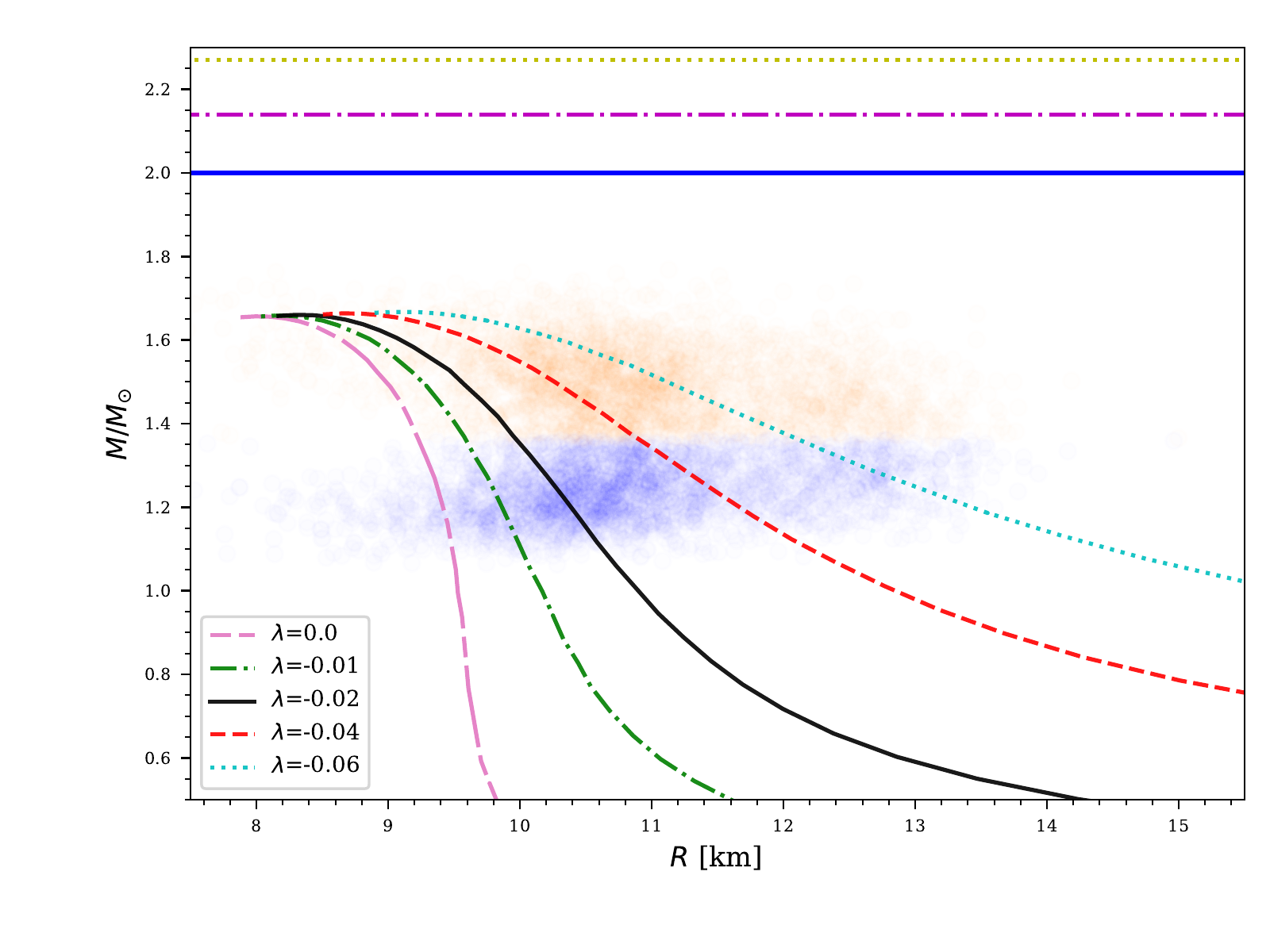}
    \caption{APR1 equation of state.}\label{fig:mr_apr1}
  \end{subfigure}
  \begin{subfigure}[b]{0.495\textwidth}
    \includegraphics[scale=0.47]{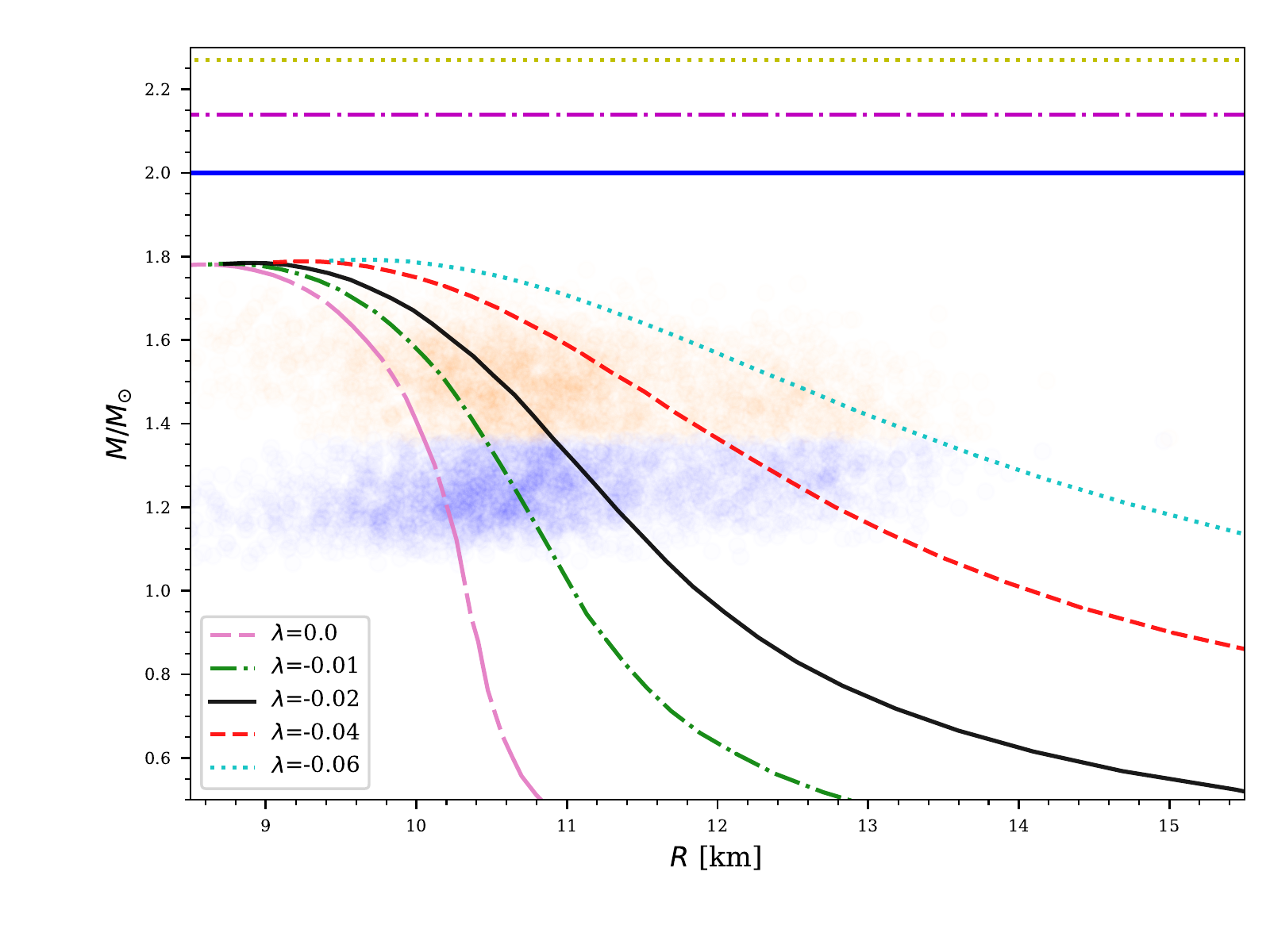}
    \caption{APR2 equation of state.}\label{fig:mr_apr2}
  \end{subfigure}
\begin{subfigure}[b]{0.495\textwidth}
     \includegraphics[scale=0.47]{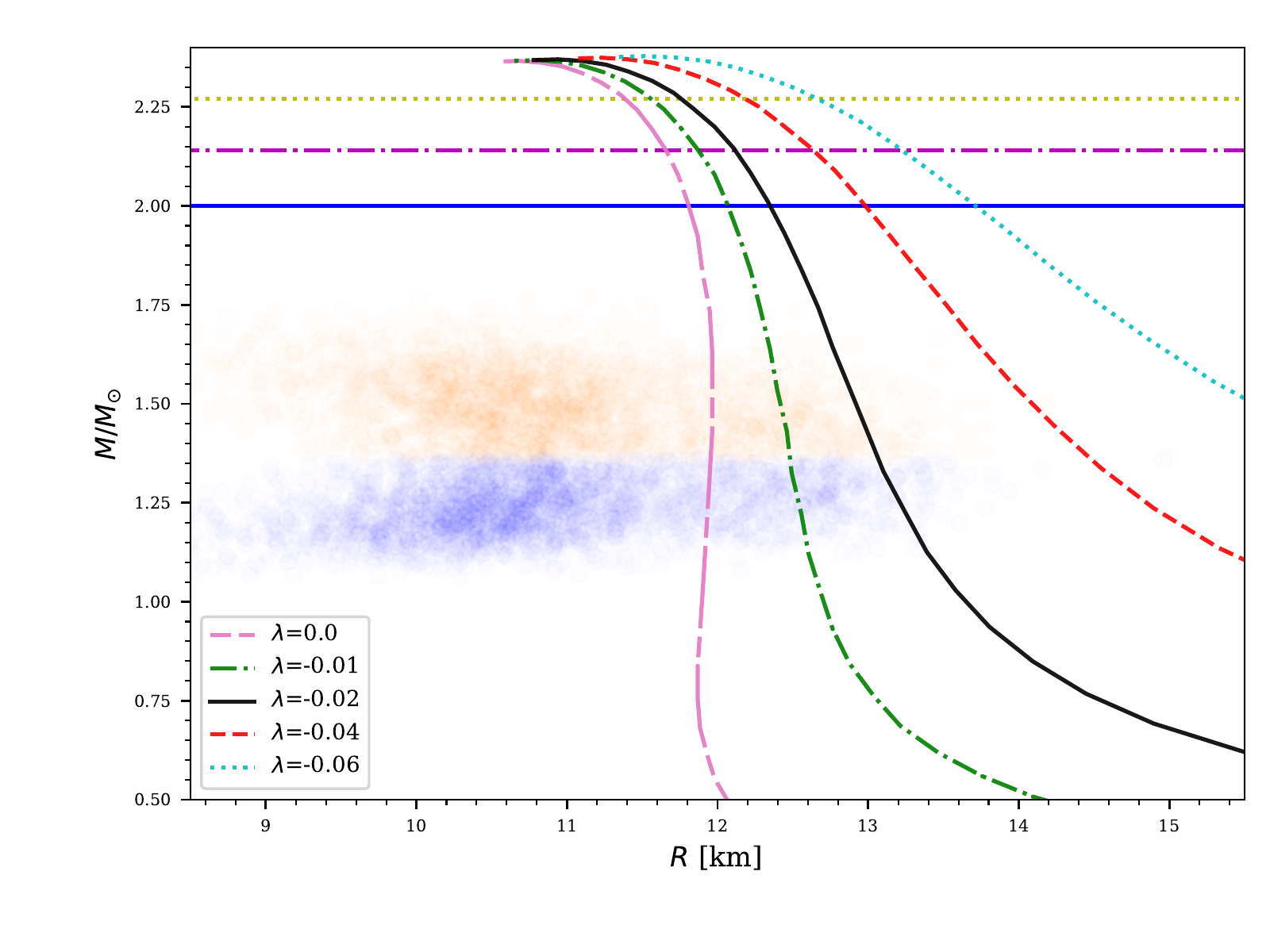}
      \caption{APR3 equation of state.}\label{fig:mr_APR3}
    \end{subfigure}
    \begin{subfigure}[b]{0.495\textwidth}
    \includegraphics[scale=0.47]{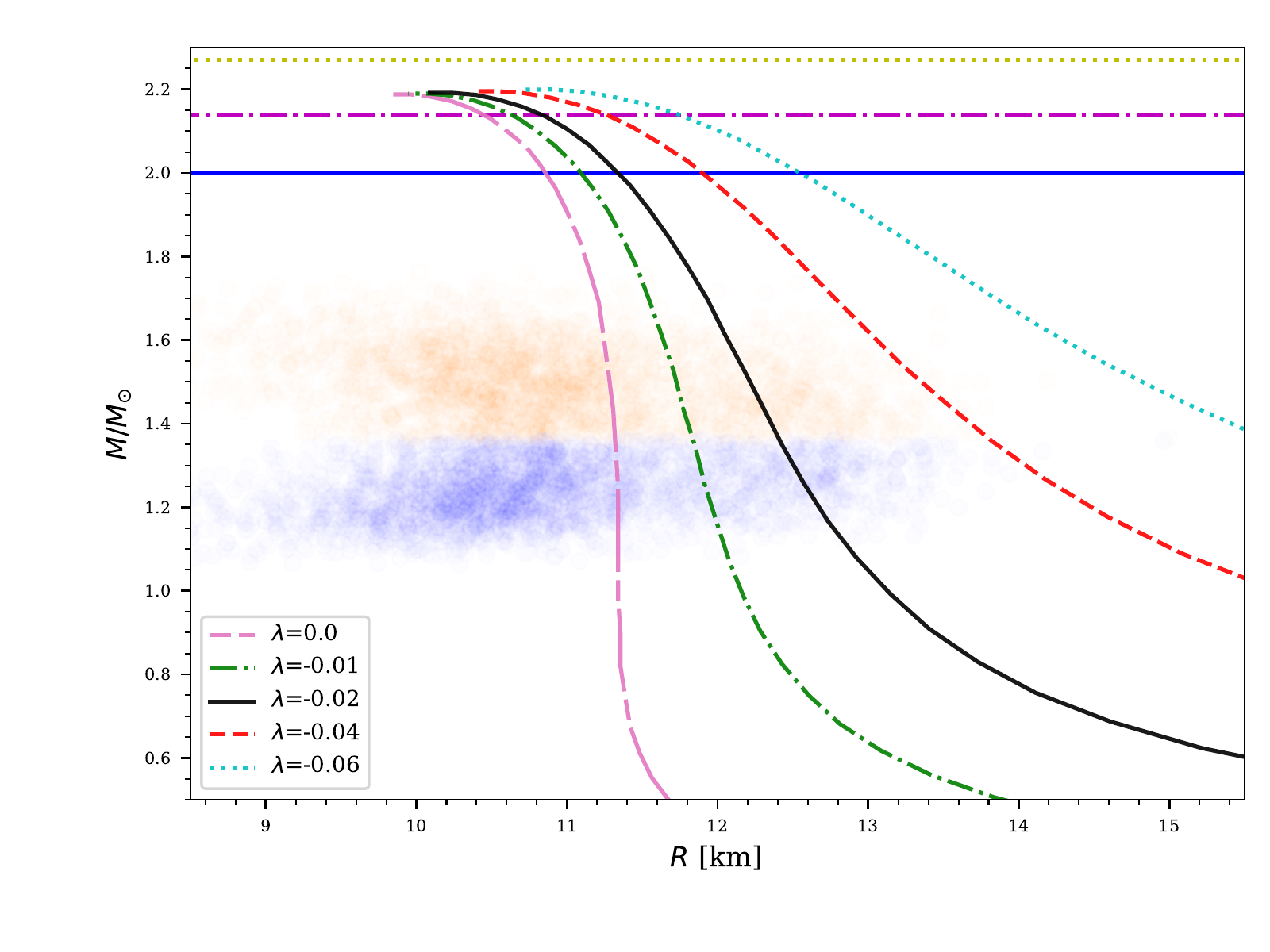}
    \caption{APR4 equation of state.}\label{fig:mr_APR4}
  \end{subfigure}
  \caption{Mass-radius relation: On the upper left side, the mass-radius relation for the APR1 equation
    of state. On the upper right side, the mass-radius relation for the APR2 equation of state. On
    the lower left side, the mass-radius relation for the APR3 equation
    of state. On the lower right side, the mass-radius relation for the APR4 equation of state. It was
    considered five values of $\lambda$ in the mass-radius for each EoS, going from $\lambda=-0.06$ to 0.0, for $\lambda=0$, the theory retrieves general relativity. The blue and orange cloud region is the constraints for
      mass-radius from the GW170817 event, which was a merger of two neutron stars with an
      observation in the electromagnetic and gravitational spectrum. The blue continuous line at
      2.0~$M{\odot}$, the magenta dot-dashed line at 2.14~$M{\odot}$ and the yellow dot line at
      2.27~$M{\odot}$ represent the most massive
      pulsars observed up to now. The pulsar with 2.14~$M{\odot}$ has a 95.4\% credibility level.}\label{fig:mr_apr}
  \end{figure}

In Figure~\eqref{fig:mr_apr}, we present the mass-radius relation for all parametrizations of the APR equation of state; this EoS considers modern nuclear potentials and frequently appears in the literature. One of its parametrization, the APR4, was constrained within the
LIGO-VIRGO~\cite{theligoscientificcollaborationandthevirgocollaboration/2018,
  ligoscientificcollaborationandvirgocollaboration/2019} observational area, highlighted as the blue
and orange cloud regions in the figures. The top orange region corresponds to the heavier and the bottom
blue region to the lighter NS detected.

We generated mass-radius curves for all sets of parametrizations, and for five different
values of $\lambda$, being $\lambda=0$ the curves for general relativity and $\lambda=-0.06, -0.04,
-0.02$ and $-0.01$ the curves for the $f(\mathcal{R,T})$ gravity. We also plotted a continuous blue line representing the 2.0~$M{\odot}$ pulsar, a magenta dot-dashed line representing the 2.14~$M{\odot}$ pulsar, and a yellow dotted line representing the 2.27~$M{\odot}$ pulsar. They represent the most massive pulsars observed up to now.

On the upper side of Figure~\eqref{fig:mr_apr} it is possible to see the first two parametrizations, APR1 Figure~\eqref{fig:mr_apr1}, and APR2 Figure~\eqref{fig:mr_apr2}. If one considers GR, i.e., $\lambda=0$, the
curves give a radius of less than 10~km for stars with more than 0.6 solar mass; this makes stars with
$M > 1.4~M_{\odot}$ out of cloud region by the GW170817 event. Considering the contributions
from the $f(\mathcal{R,T})$ gravity, it is possible to see an increase in the radius in both cases, and the curves
are brought inside to the cloud region. In Figure~\eqref{fig:mr_apr1}, the curves between $\lambda=-0.04$
and $-0.02$ present values of mass-radius better ranged within the
observational region constrained by the LIGO-VIRGO collaboration. In the case of Figure~\eqref{fig:mr_apr2}, values
between $-0.02$ and $-0.01$ are the better ones. We can observe that for the APR1 and APR2 EsoS,
contributions from the $f(\mathcal{R,T})$ gravity can put the curves within the
LIGO-VIRGO constrains for the radius; however, it is not possible to reach two solar mass with these
two parametrizations.

On the lower side of Figure~\eqref{fig:mr_apr}, we have the APR3~\eqref{fig:mr_APR3} and APR4~\eqref{fig:mr_APR4}
parametrizations. In the case of GR, it is possible to see that the curves are inside the
LIGO-VIRGO region. For APR3 we can see that a 2.27~$M_{\odot}$ pulsar could be
explained by this parametrization, although there is substantial uncertainty in the mass of this
one. The two parametrizations could reach the massive NS with 2.14~$M_{\odot}$ and satisfying the
mass-radius region of the LIGO-VIRGO observation simultaneously. In both Figures, one can see that the $\lambda$'s
minimum value is around $-0.02$; otherwise, the mass-radius starts to stay out of the
cloud region. The LIGO-VIRGO teams well studied the APR4 parametrization. The mass-radius and tidal parameters are the most promising parametrization of this EoS. However, according to Radice et al.~\cite{radice/2018a}, the APR4, as well as the FPS (that we will see ahead) are tentatively excluded by the electromagnetic (EM) counterpart of the multi-messenger observation.

\begin{figure}[ht]
    \centering
  \begin{subfigure}[b]{0.495\textwidth}
    \includegraphics[scale=0.47]{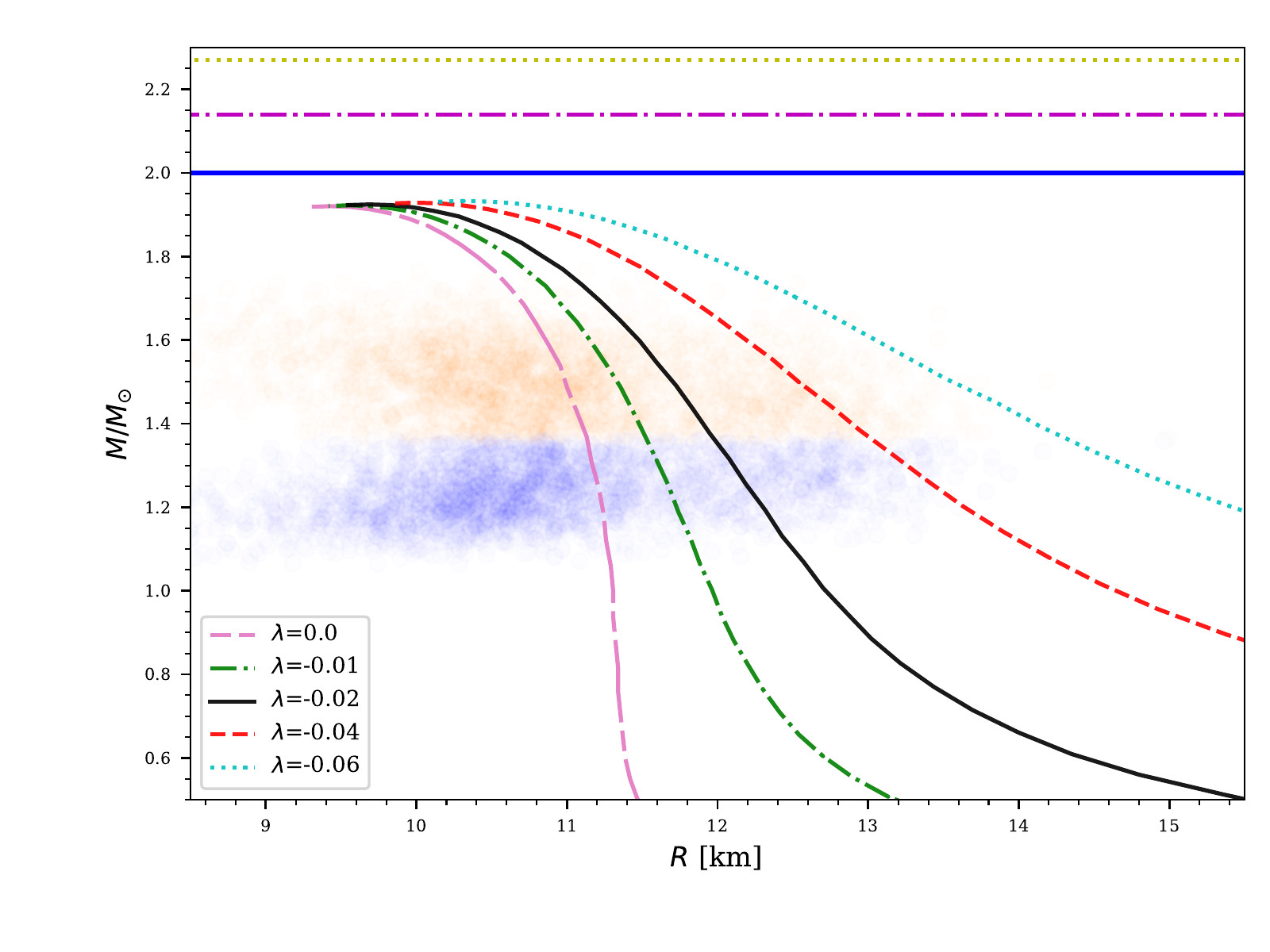}
    \caption{BBB2 equation of state.}\label{fig:mr_bbb2}
  \end{subfigure}
  \begin{subfigure}[b]{0.495\textwidth}
      \includegraphics[scale=0.47]{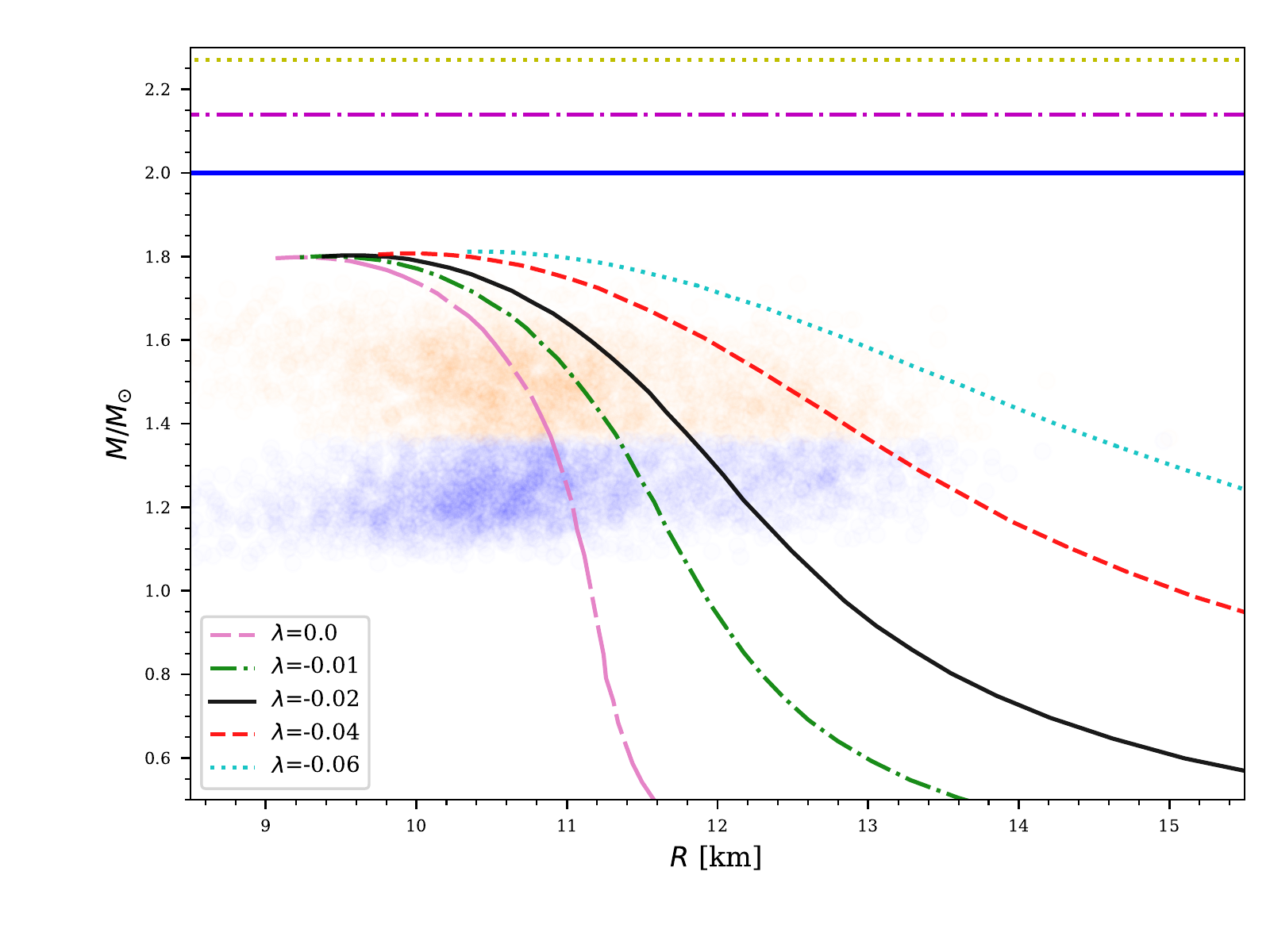}
      \caption{FPS equation of state.}\label{fig:mr_fps}
    \end{subfigure}
  \caption{Mass-radius relation: On the left side the mass-radius relation for the BBB2 equation
    of state. On the right side the mass-radius relation for the FPS equation of state.
  It was
    considered five values of $\lambda$ in the mass-radius for each EoS, going from $\lambda=-0.06$ to 0.0, for $\lambda=0$, the theory retrieves general relativity. The blue and orange cloud region is the constraints for
      mass-radius from the GW170817 event, which was a merger of two neutron stars with an
      observation in the electromagnetic and gravitational spectrum. The blue continuous line at
      2.0~$M{\odot}$, the magenta dot-dashed line at 2.14~$M{\odot}$ and the yellow dot line at
      2.27~$M{\odot}$ represent the most massive
      pulsars observed up to now. The pulsar with 2.14~$M{\odot}$ has a 95.4\% credibility level.}\label{fig:mr_bbfps}
  \end{figure}

In Figure~\eqref{fig:mr_bbfps}, we present two EsoS;\ the BBB2, which is a parametrization of the BBB
EoS and the FPS one. The BBB2 in a non-relativistic EoS in the framework of BBG, it is for
medium-stiff EoS with crust thickness about 0.8 km~\cite{haensel/2007}. We show the BBB2 EoS in
Figure~\eqref{fig:mr_bbb2}, being the theory for GR ($\lambda=0)$ inside the LIGO-VIRGO cloud region. However, it does
not achieve two solar masses. The $f(\mathcal{R,T})$ gravity theory increases a small fraction of the mass, but it is not
enough to reach significant values; the radius, on the other hand, has a significant increment; the minimum value of its parameters is
around $\lambda=-0.04$. In Figure~\eqref{fig:mr_fps}, we show the mass-radius for the FPS equation
of state, and, as the previous one, it is inside the LIGO-VIRGO cloud region for GR and does not reach
the two solar mass. The minimum value of the $f(\mathcal{R,T})$ parameter is about the same also, $-0.04$. These two EsoS
can be excluded in our case since they predict masses smaller than 2~$M_{\odot}$ under GR or
$f(\mathcal{R,T})$ gravity.

\begin{figure}[ht]
    \centering
    \begin{subfigure}[b]{0.495\textwidth}
      \includegraphics[scale=0.47]{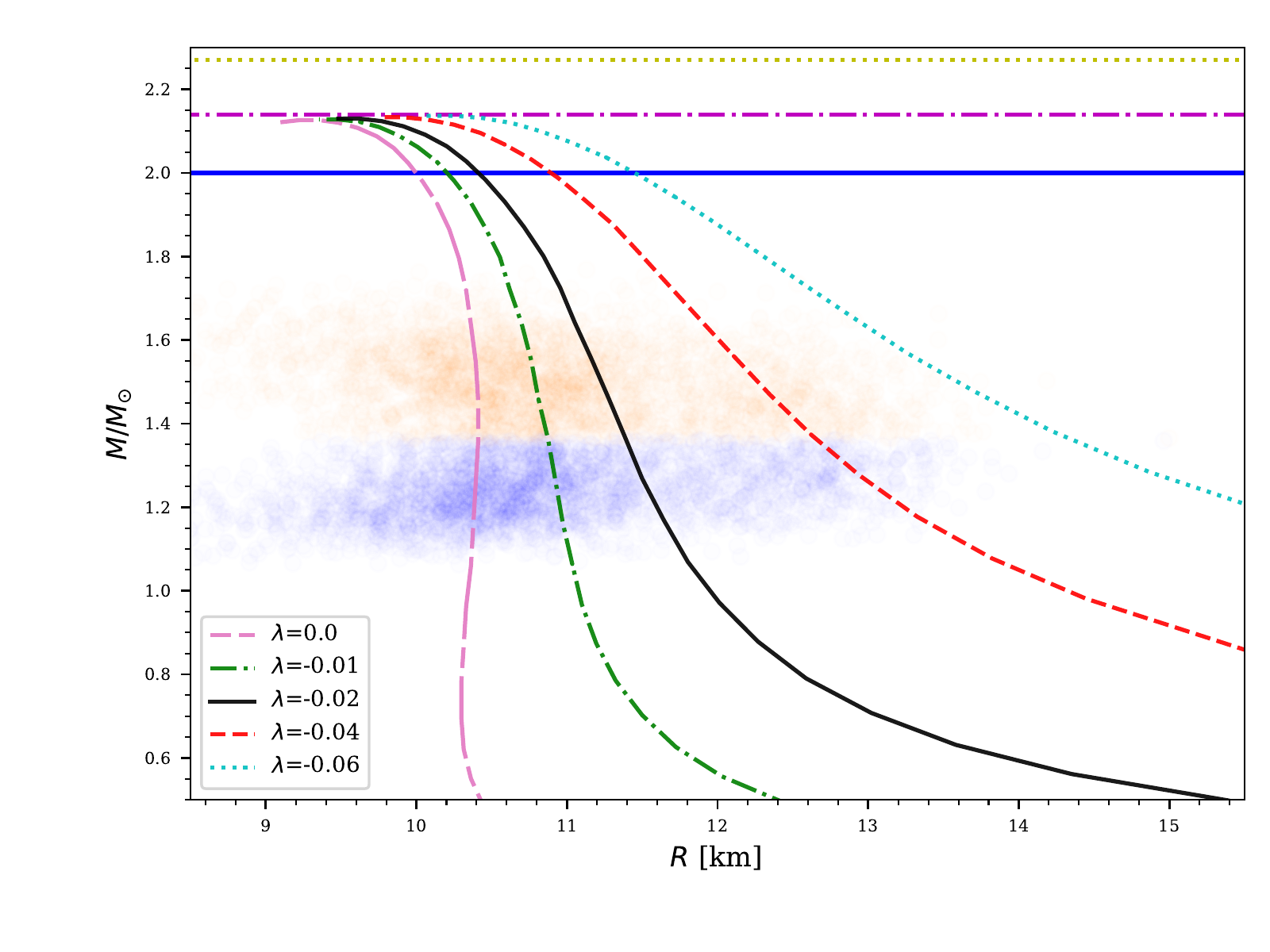}
      \caption{WFF1 equation of state.}\label{fig:mr_wff1}
  \end{subfigure}
  \begin{subfigure}[b]{0.495\textwidth}
    \includegraphics[scale=0.47]{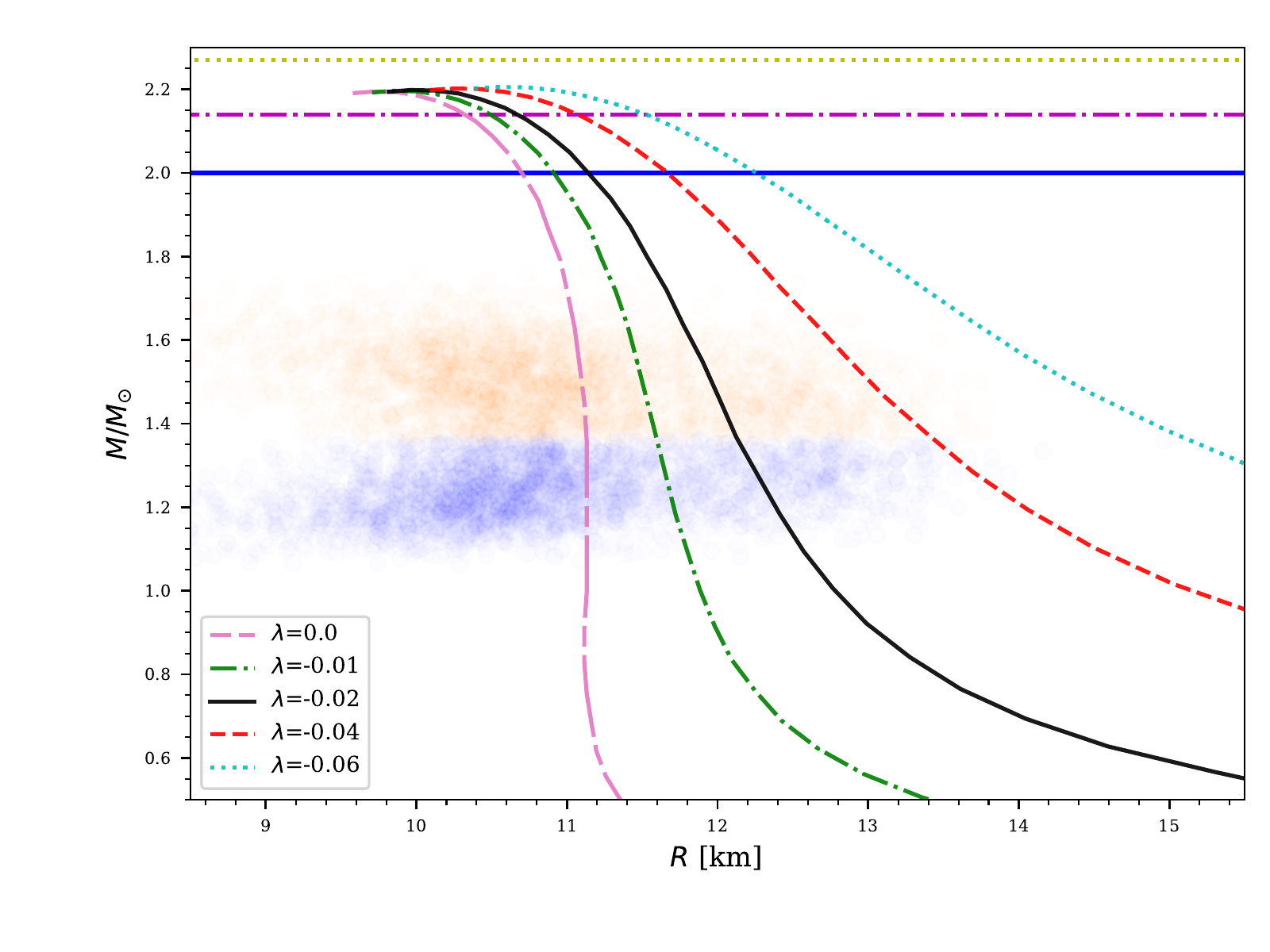}
    \caption{WFF2 equation of state.}\label{fig:mr_wff2}
  \end{subfigure}
      \begin{subfigure}[b]{0.495\textwidth}
      \includegraphics[scale=0.47]{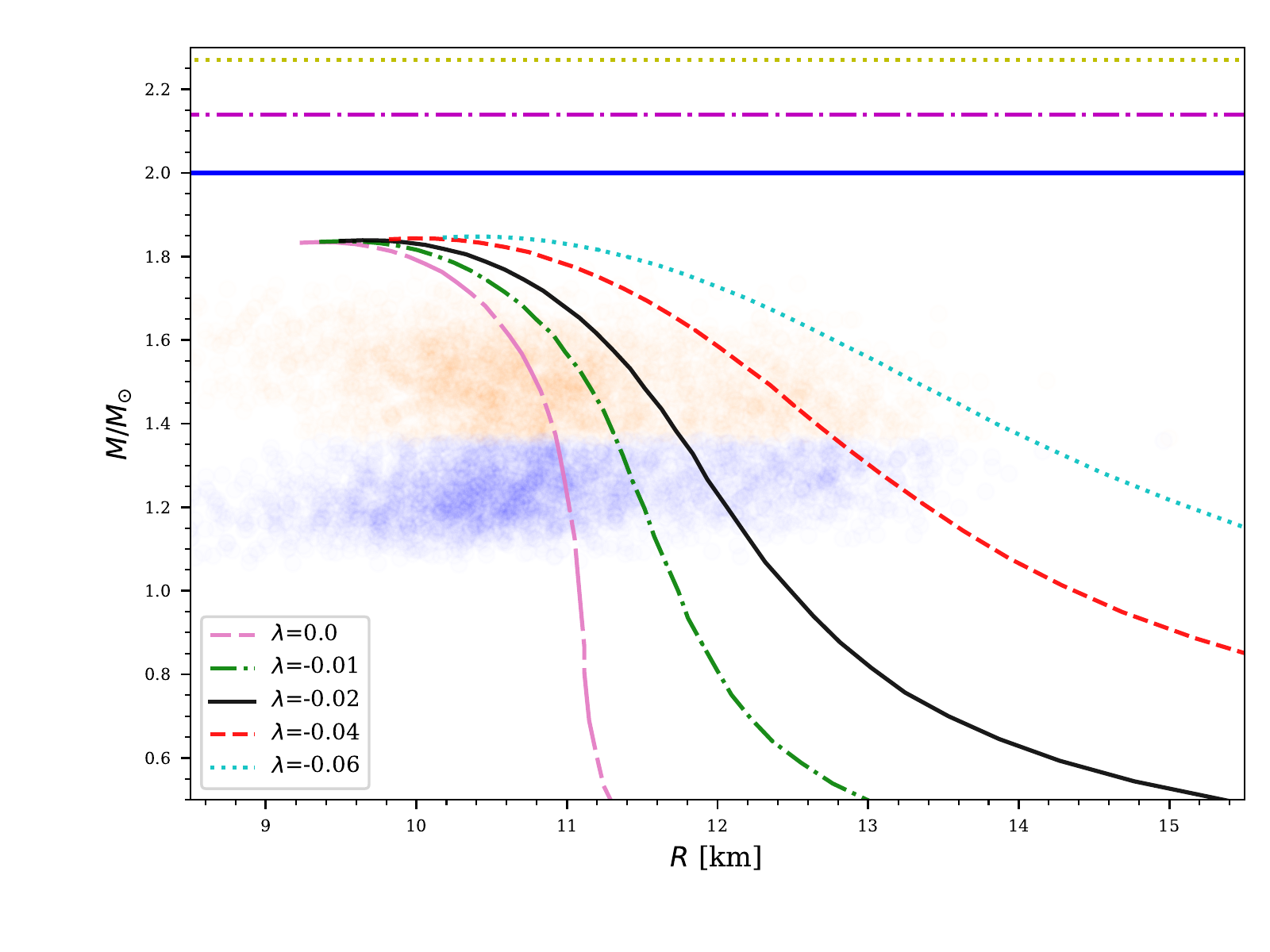}
      \caption{WFF3 equation of state.}\label{fig:mr_wff3}
  \end{subfigure}
  \caption{Mass-radius relation: Upper on the left side the mass-radius relation for the WFF1 equation
    of state. Upper on the right side the mass-radius relation for the WFF2 equation of state. Lower
    the mass-radius relation for the WFF3 quation of state.
It was
    considered five values of $\lambda$ in the mass-radius for each EoS, going from $\lambda=-0.06$ to 0.0, for $\lambda=0$, the theory retrieves general relativity. The blue and orange cloud region is the constraints for
      mass-radius from the GW170817 event, which was a merger of two neutron stars with an
      observation in the electromagnetic and gravitational spectrum. The blue continuous line at
      2.0~$M{\odot}$, the magenta dot-dashed line at 2.14~$M{\odot}$ and the yellow dot line at
      2.27~$M{\odot}$ represent the most massive
      pulsars observed up to now. The pulsar with 2.14~$M{\odot}$ has a 95.4\% credibility level.}\label{fig:mr_wff}
  \end{figure}

In Fig.~\eqref{fig:mr_wff}, we present the mass-radius relationship for the WFF EoS and its set of parametrizations. On the upper side, we have two parametrizations, WFF1~\eqref{fig:mr_wff1} and
WFF2~\eqref{fig:mr_wff2}, on the left and right sides, respectively. Both sets reach more than two
solar masses and are inside the cloud region considering GR;~the lower limit of the $f(\mathcal{R,T})$
parameter is around $-0.04$ for both cases. In Figure~\eqref{fig:mr_wff1}, the maximum mass is near 2.14~$M_{\odot}$. We can see that the $f(\mathcal{R,T})$ increases a bit the maximum mass.
However, one can go to higher values of $\lambda$ in modulus. The LIGO-VIRGO collaborations' paper constrained the parametrization WFF1. Using tidal deformability through GW and EM measurements, Coughlin et al.~\cite{coughlin/2018}
have disfavored WFF1 compared with other EsoS. In Figure~\eqref{fig:mr_wff2}, we show the WFF2 EoS, and, as in the previous case, it is possible to
reach 2.14~$M_{\odot}$. This parametrization allows an increase in the mass and
radius compared to WFF1. In Figure~\eqref{fig:mr_wff3}, we present the WFF3 parametrization, which has an intermediate stiffness. As we can see, the gravitational mass of this parametrization does not reach more than two solar mass, either in GR or $f(\mathcal{R,T})$ gravity.

\begin{figure}[ht]
    \centering
    \begin{subfigure}[b]{0.495\textwidth}
      \includegraphics[scale=0.47]{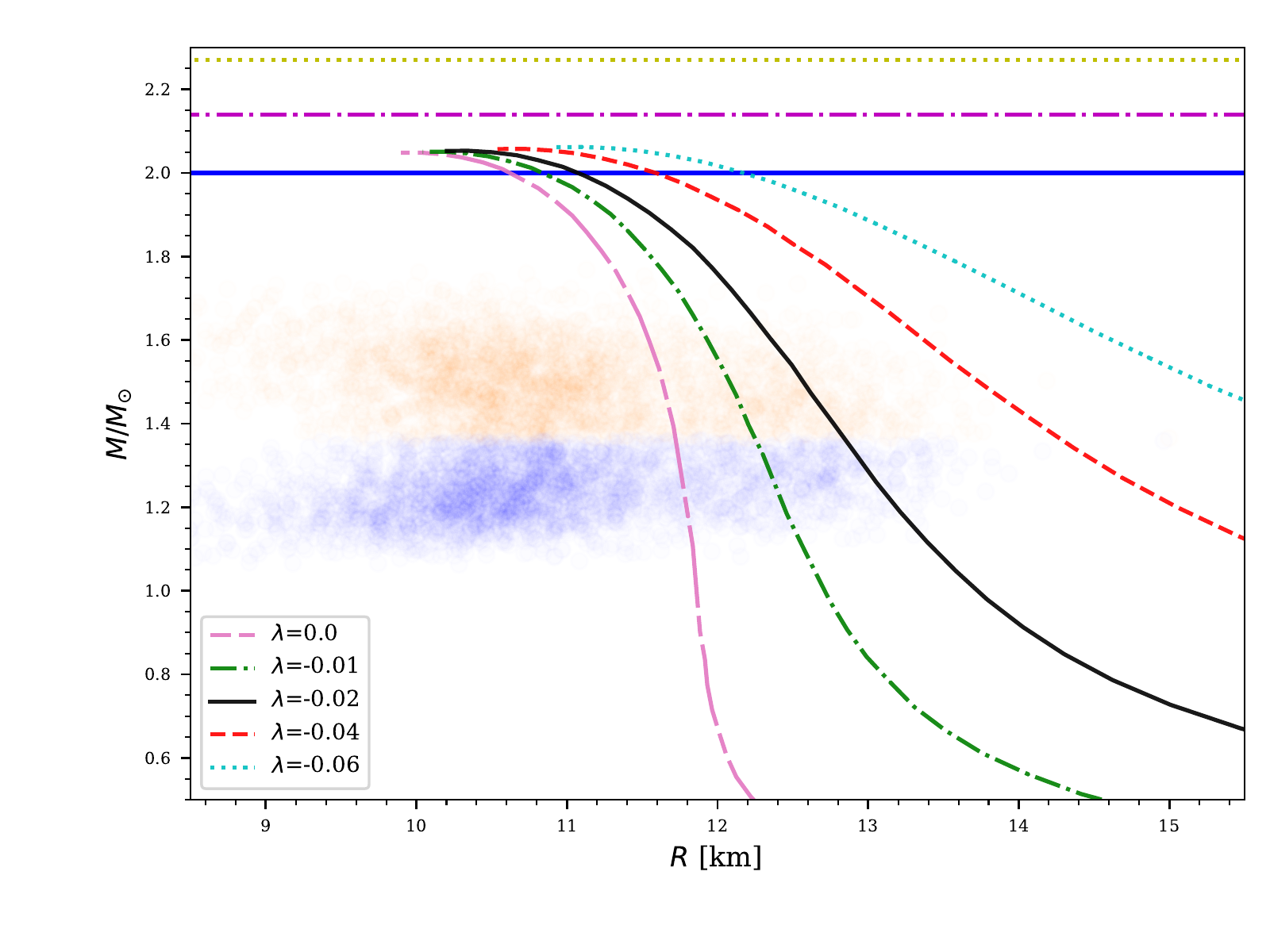}
      \caption{SLy equation of state.}\label{fig:mr_sly}
    \end{subfigure}
    \begin{subfigure}[b]{0.495\textwidth}
      \includegraphics[scale=0.47]{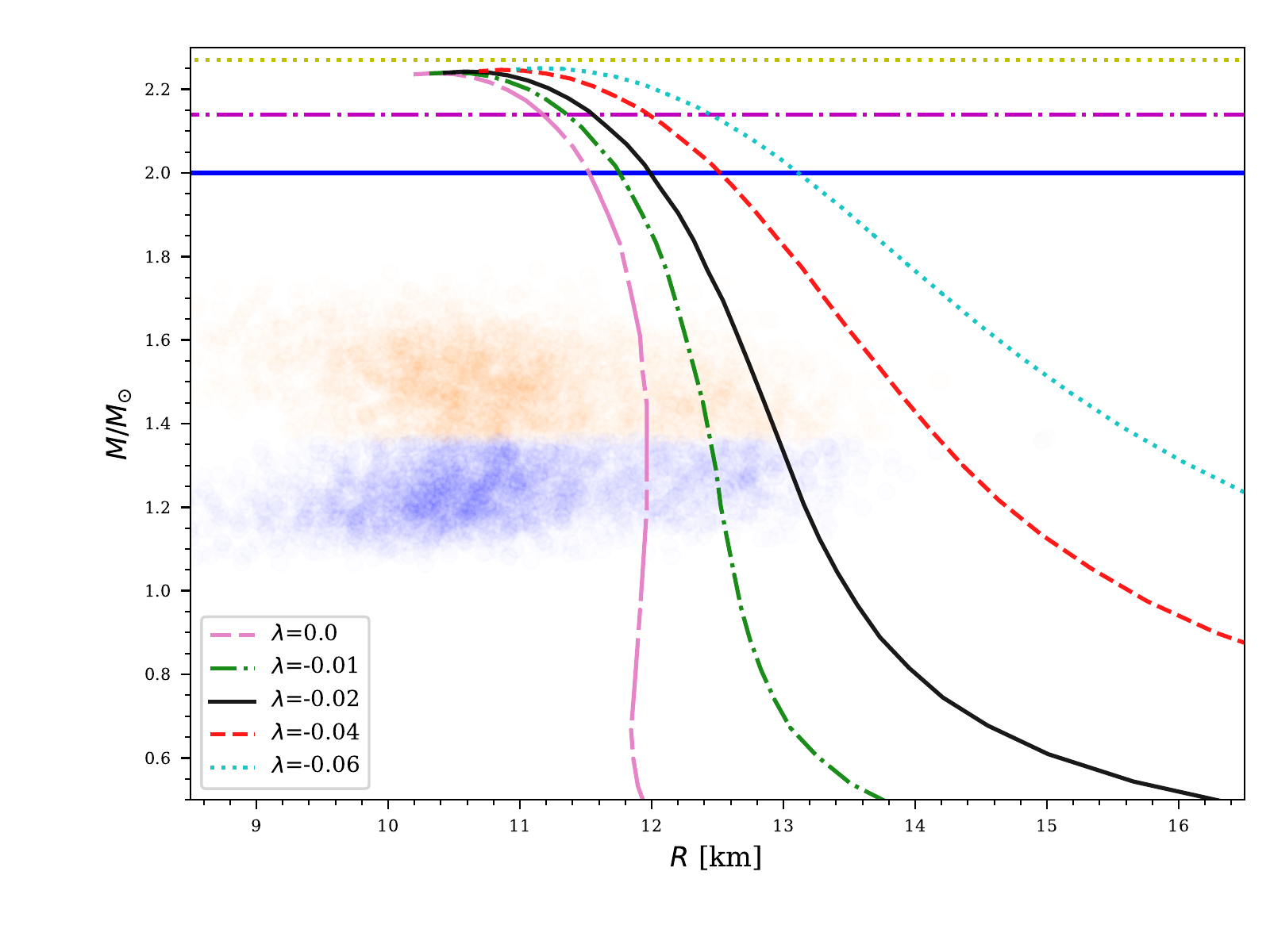}
      \caption{ENG equation of state.}\label{fig:mr_eng}
    \end{subfigure}
    \caption{Mass-radius relation: On the left side the mass-radius relation for the SLy equation
      of state. On the right side the mass-radius relation for the ENG equation of state. It was
    considered five values of $\lambda$ in the mass-radius for each EoS, going from $\lambda=-0.06$ to 0.0, for $\lambda=0$, the theory retrieves general relativity. The blue and orange cloud region is the constraints for
      mass-radius from the GW170817 event, which was a merger of two neutron stars with an
      observation in the electromagnetic and gravitational spectrum. The blue continuous line at
      2.0~$M{\odot}$, the magenta dot-dashed line at 2.14~$M{\odot}$ and the yellow dot line at
      2.27~$M{\odot}$ represent the most massive
      pulsars observed up to now. The pulsar with 2.14~$M{\odot}$ has a 95.4\% credibility level.}\label{fig:mr_sleng}
  \end{figure}

In Figure~\eqref{fig:mr_sleng}, we have two equations of state, the SLy on the left
side, Figure~\eqref{fig:mr_sly}, and the ENG on the right side, Figure~\eqref{fig:mr_eng}. The SLy is a
well-studied EoS either in modified theories (with analytical representation in
$f(R)$, for example) or General Relativity. It is a Skyrme type
EoS with an effective nuclear interaction and was considered in the LIGO-VIRGO work. As one can see,
the SLy can reach two solar masses and is inside the mass-radius cloud region from
LIGO-VIRGO observation. However, it cannot achieve 2.14~$M_{\odot}$. This EoS is often used to describe the star's inner crust, while the BPS~\cite{baym/1971a} EoS describes the outer crust. In GR, the crustal EsoS have little importance to the global parameters of neutron stars~\cite{haensel/2007}.

The SLy describes the core and inner crust in a unified manner. If we consider the $f(\mathcal{R,T})$
gravity, the minimum value for the $\lambda$ parameter is around $-0.02$, similar to APR3--4. On
the right side of Figure~\eqref{fig:mr_sleng}, we have the ENG equation of state,
Figure~\eqref{fig:mr_eng}; it is an EoS that reaches 2.14~$M_{\odot}$ and is inside the cloud
region if one considers general relativity. This EoS was studied in an entirely realistic hydrodynamic simulation~\cite{east/2019},
considering spin and other parameters in the context of GW170817, and this context of binary star
merger if it could lead to the formation of a supra massive NS~\cite{ma/2018}. Using neural networks,
Fujimoto et al.~\cite{fujimoto/2020} showed that ENG and other many-body models, are
favoured. Considering the $f(\mathcal{R,T})$ gravity, this EoS has similar behaviour with the APR3--4 and SLy,
i.e., the lower value of $\lambda$ is $-0.02$ for the mass-radius to stay inside the cloud
region observed. The $f(\mathcal{R,T})$ gravity increases the mass, however not enough to reach a 2.27~$M_{\odot}$, for
example.

\begin{figure}[ht]
    \centering
    \begin{subfigure}[b]{0.495\textwidth}
      \includegraphics[scale=0.47]{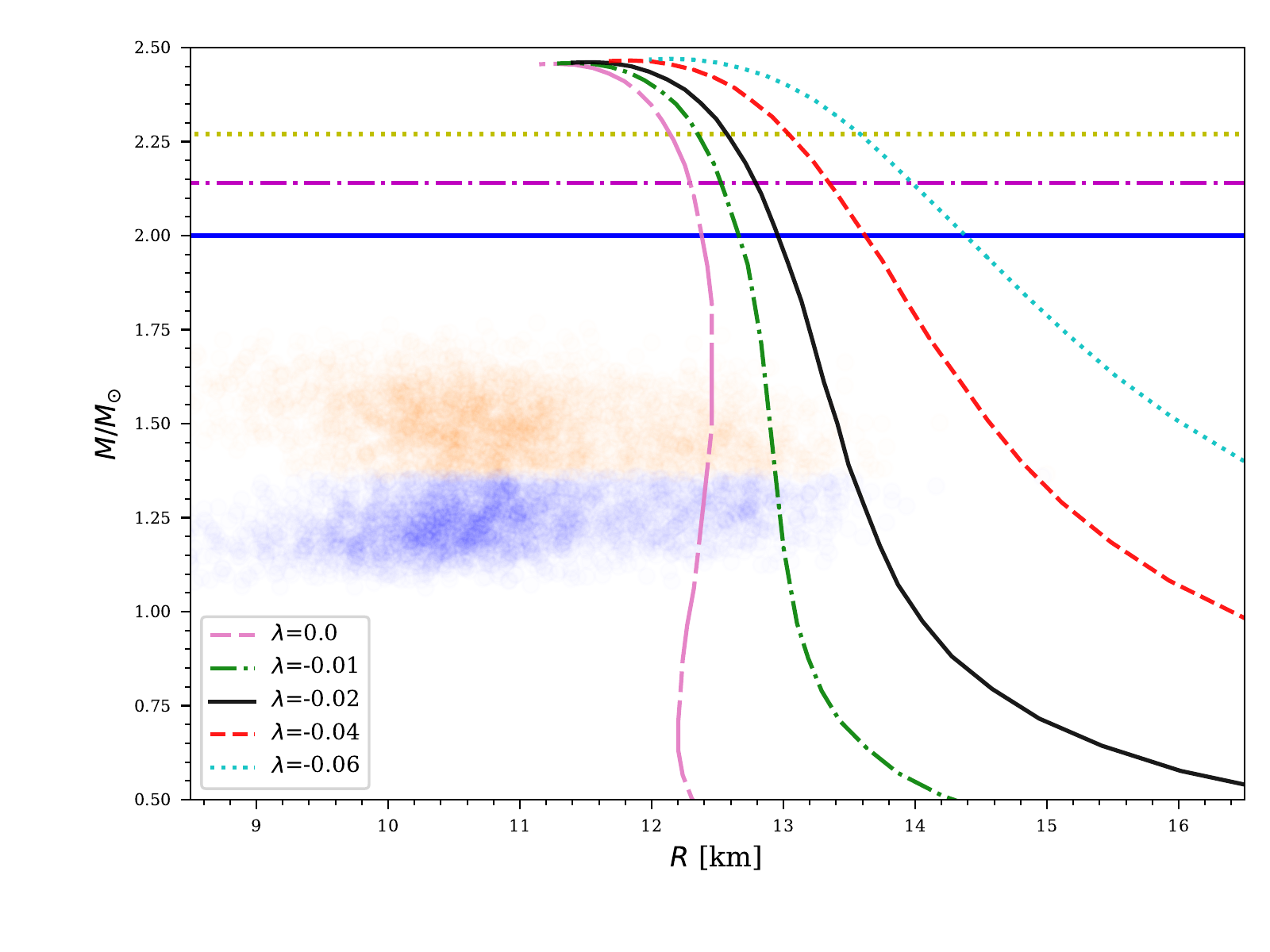}
      \caption{MPA1 equation of state.}\label{fig:mr_MPA1}
    \end{subfigure}
    \caption{Mass-radius relation for the MPA1 equation
      of state
    It was
    considered five values of $\lambda$ in the mass-radius for each EoS, going from $\lambda=-0.06$ to 0.0, for $\lambda=0$, the theory retrieves general relativity. The blue and orange cloud region is the constraints for
      mass-radius from the GW170817 event, which was a merger of two neutron stars with an
      observation in the electromagnetic and gravitational spectrum. The blue continuous line at
      2.0~$M{\odot}$, the magenta dot-dashed line at 2.14~$M{\odot}$ and the yellow dot line at
      2.27~$M{\odot}$ represent the most massive
      pulsars observed up to now. The pulsar with 2.14~$M{\odot}$ has a 95.4\% credibility level.}\label{fig:mr_MP}
  \end{figure}

In Figure~\eqref{fig:mr_MP}, we show the MPA1 equation of state. This EoS was also studied in the
LIGO-VIRGO work, along with SLy, WFF1--2, ENG, and APR3--4. The MPA1 EoS can reach the two solar mass
limit if we consider only general relativity. Given the stiffness of this EoS, it is possible to reach
mass around the 2.5 value; however according to Ma et al.~\cite{ma/2018}, the absence of a supra
massive NS signature in the event GW170817/AT2017gfo could rule out the MPA1 or even the APR3. The
lower boundary for the $\lambda$ parameter is around $-0.02$ as the other stiff EsoS.

\begin{figure}[ht]
    \centering
    \begin{subfigure}[b]{0.495\textwidth}
    \includegraphics[scale=0.47]{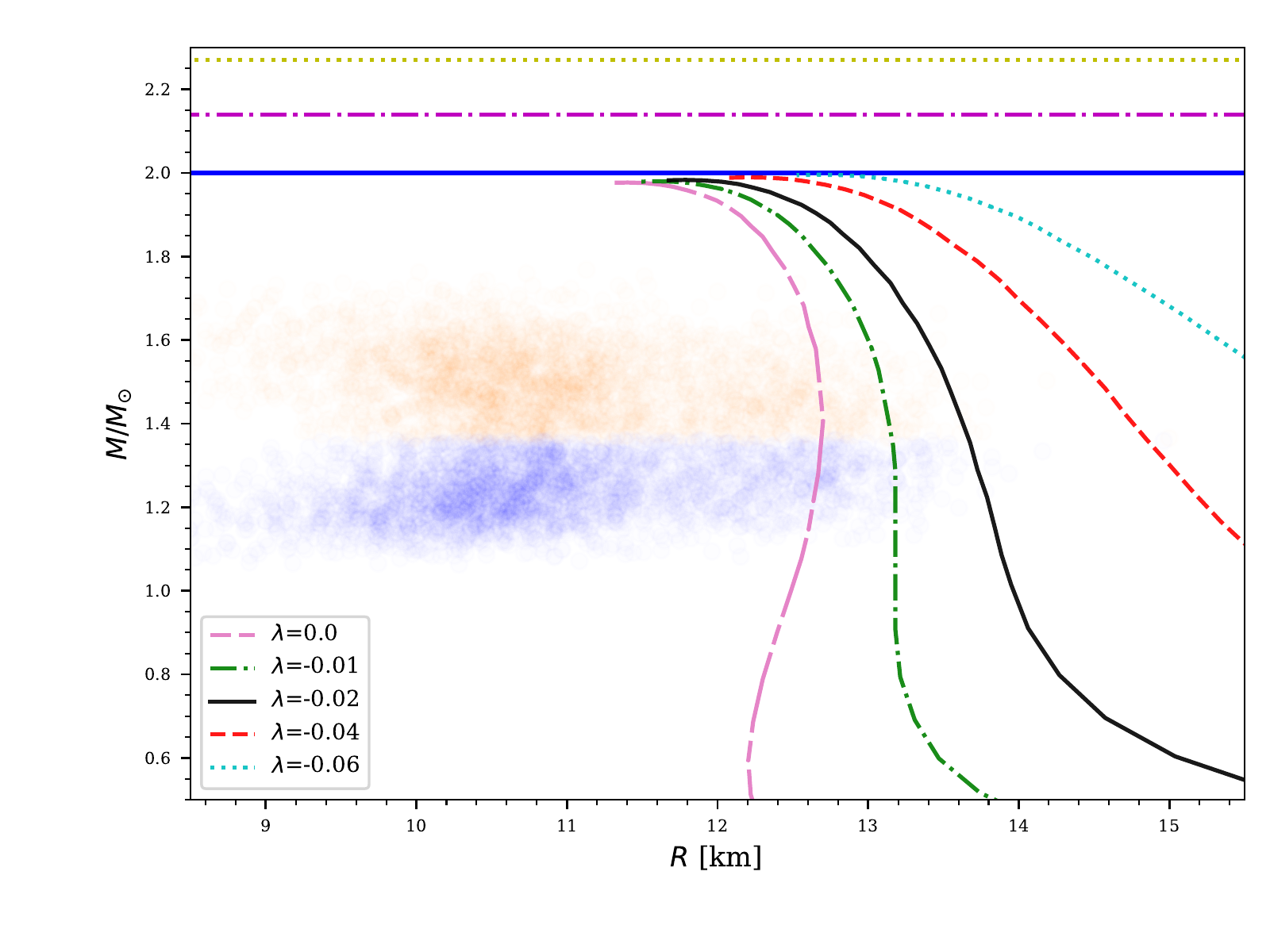}
      \caption{ALF2 equation of state.}\label{fig:mr_alf2}
    \end{subfigure}
    \begin{subfigure}[b]{0.495\textwidth}
    \includegraphics[scale=0.47]{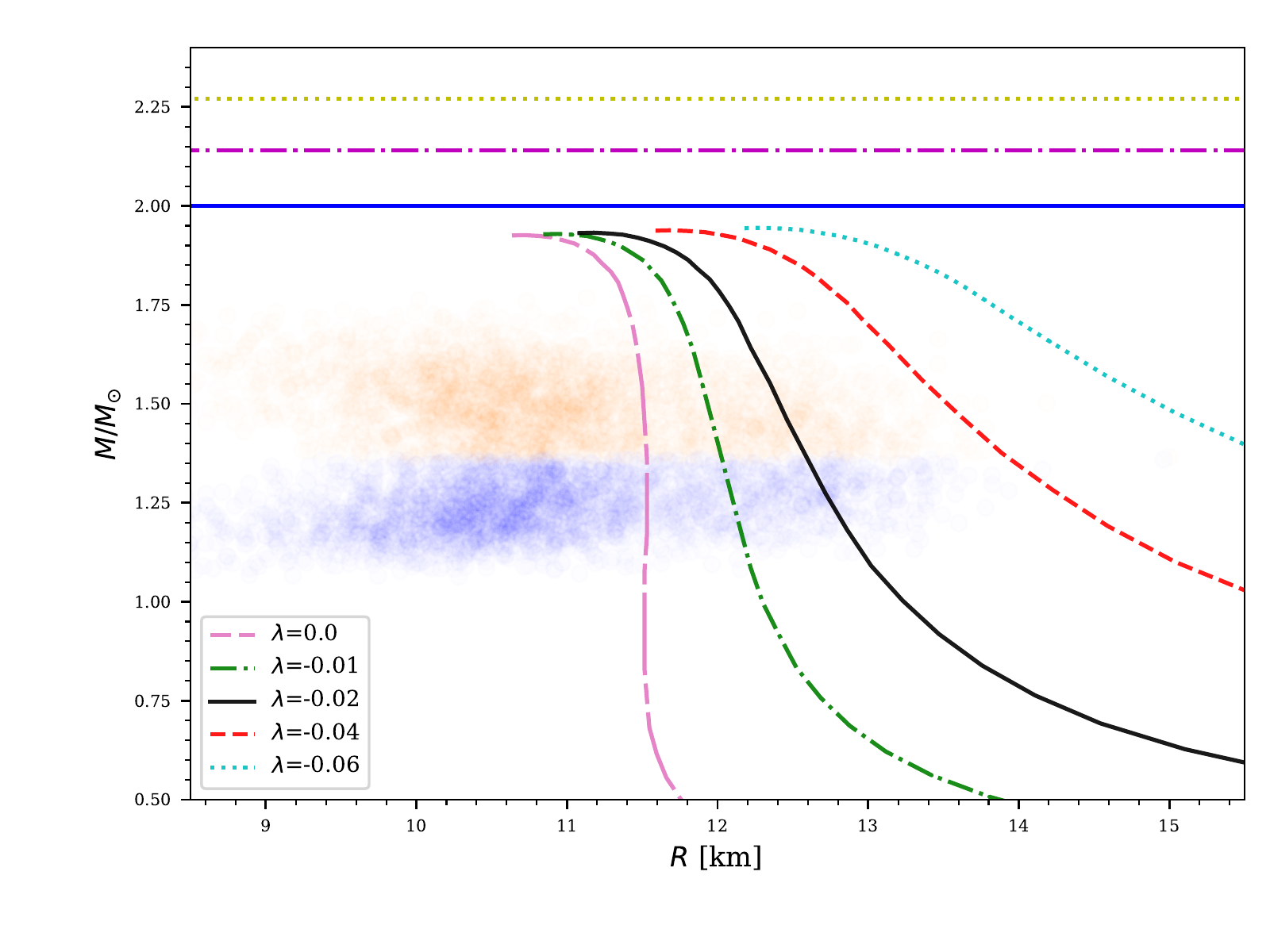}
      \caption{ALF4 equation of state.}\label{fig:mr_alf4}
    \end{subfigure}
    \caption{Mass-radius relation: On the left side the mass-radius relation for the ALF2 equation
      of state. On the right side the mass-radius relation for the ALF4 equation of state. It was
    considered five values of $\lambda$ in the mass-radius for each EoS, going from $\lambda=-0.06$ to 0.0, for $\lambda=0$, the theory retrieves general relativity. The blue and orange cloud region is the constraints for
      mass-radius from the GW170817 event, which was a merger of two neutron stars with an
      observation in the electromagnetic and gravitational spectrum. The blue continuous line at
      2.0~$M{\odot}$, the magenta dot-dashed line at 2.14~$M{\odot}$ and the yellow dot line at
      2.27~$M{\odot}$ represent the most massive
      pulsars observed up to now. The pulsar with 2.14~$M{\odot}$ has a 95.4\% credibility level.}\label{fig:mr_alf}
  \end{figure}

The last EoS studied from the piecewise representation is the ALF, in Figure~\eqref{fig:mr_alf}, which leads to the possibility of hybrid stars. We have chosen two sets of parametrizations, the ALF2 in Figure~\eqref{fig:mr_alf2} and ALF4 in Figure~\eqref{fig:mr_alf4}. Other highlighted EoS is the H1--7, which includes hyperons; one of its parametrizations was constrained in the LIGO-VIRGO paper, being out of the cloud region. The H4 was also constrained by the tidal parameter in a joint constrain from multimessenger observation~\cite{radice/2018a}, being ruled out by the LIGO-VIRGO paper. Considering the ALF2, we see that this parametrization is inside the cloud region and could admit the $\lambda$ around $-0.01$ in the case of $f(\mathcal{R,T})$; however, it cannot reach $2.14~M_{\odot}$. In the case of the ALF4 parametrization, the allowed radii for $f(\mathcal{R,T})$ are broader than the previous case, and the $\lambda$ parameter could reach values less than $-0.02$. However, the maximum mass does not reach $2~M_{\odot}$ in any case.

\begin{figure}[ht]
    \centering
    \begin{subfigure}[b]{0.495\textwidth}
    \includegraphics[scale=0.47]{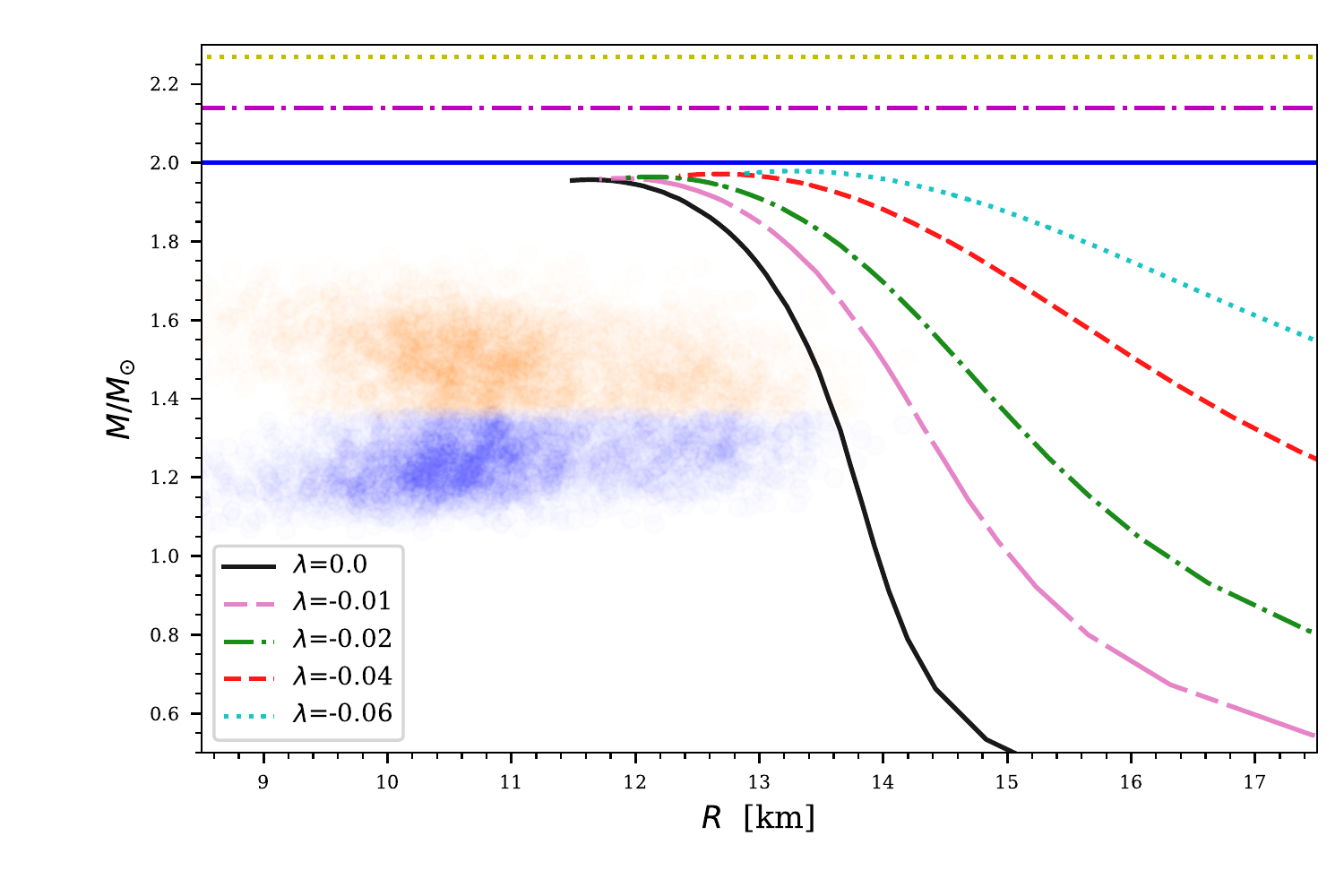}
      \caption{BKA20 equation of state.}\label{fig:mr_bka20}
    \end{subfigure}
    \begin{subfigure}[b]{0.495\textwidth}
    \includegraphics[scale=0.47]{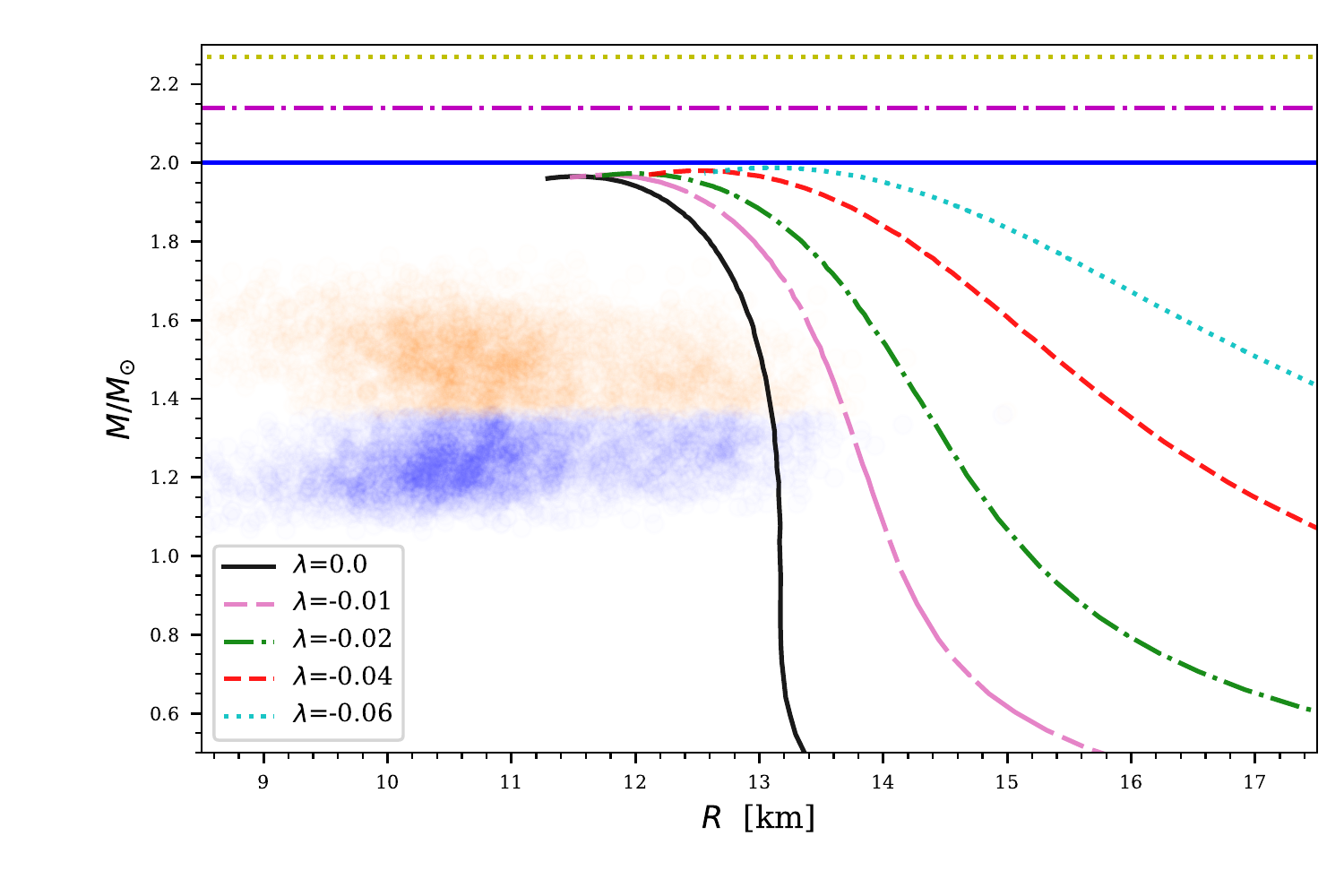}
      \caption{BSR8 equation of state.}\label{fig:mr_bsr8}
    \end{subfigure}
 \begin{subfigure}[b]{0.495\textwidth}
    \includegraphics[scale=0.47]{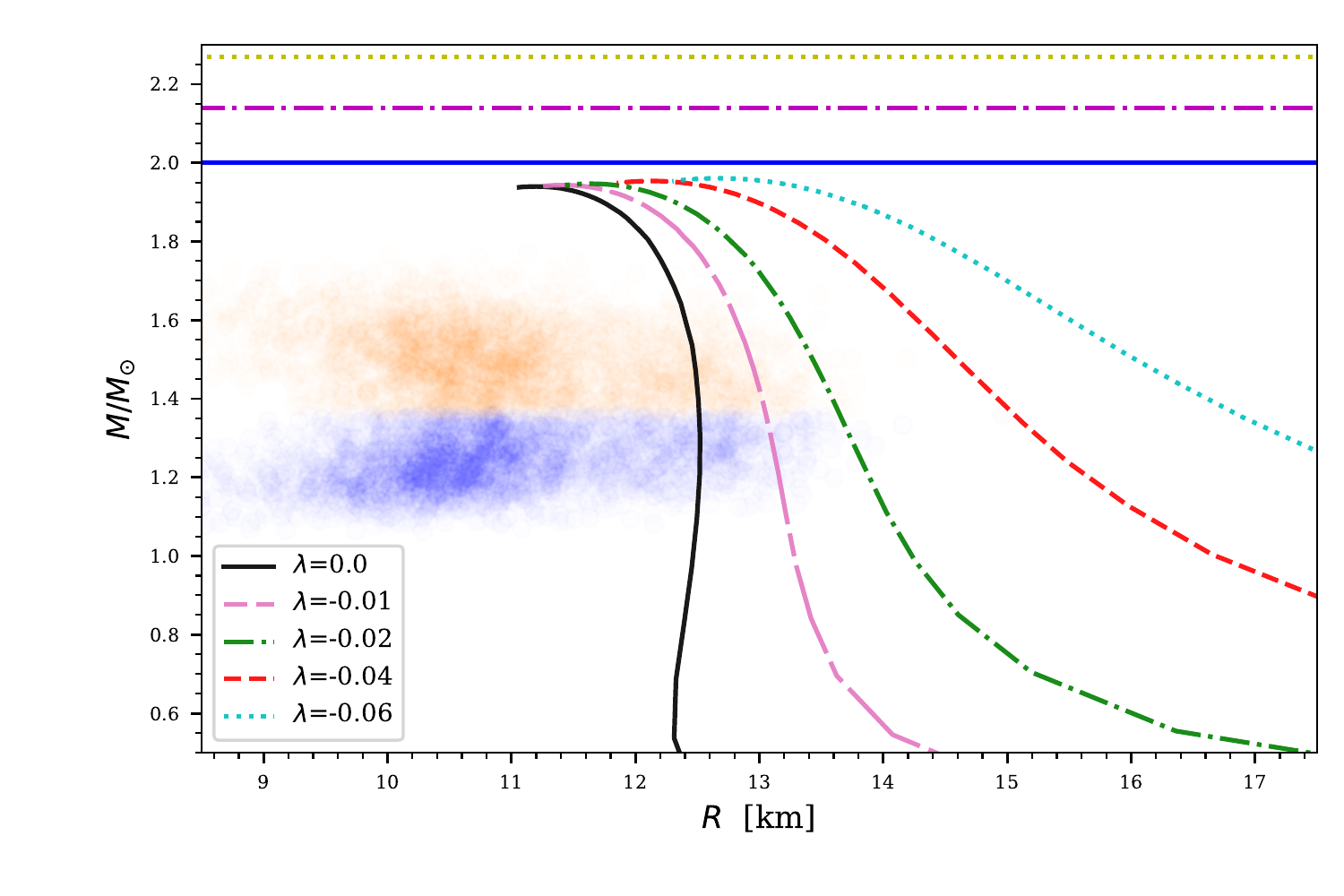}
      \caption{IU-FSU equation of state.}\label{fig:mr_iufsu}
    \end{subfigure}
    \begin{subfigure}[b]{0.495\textwidth}
    \includegraphics[scale=0.47]{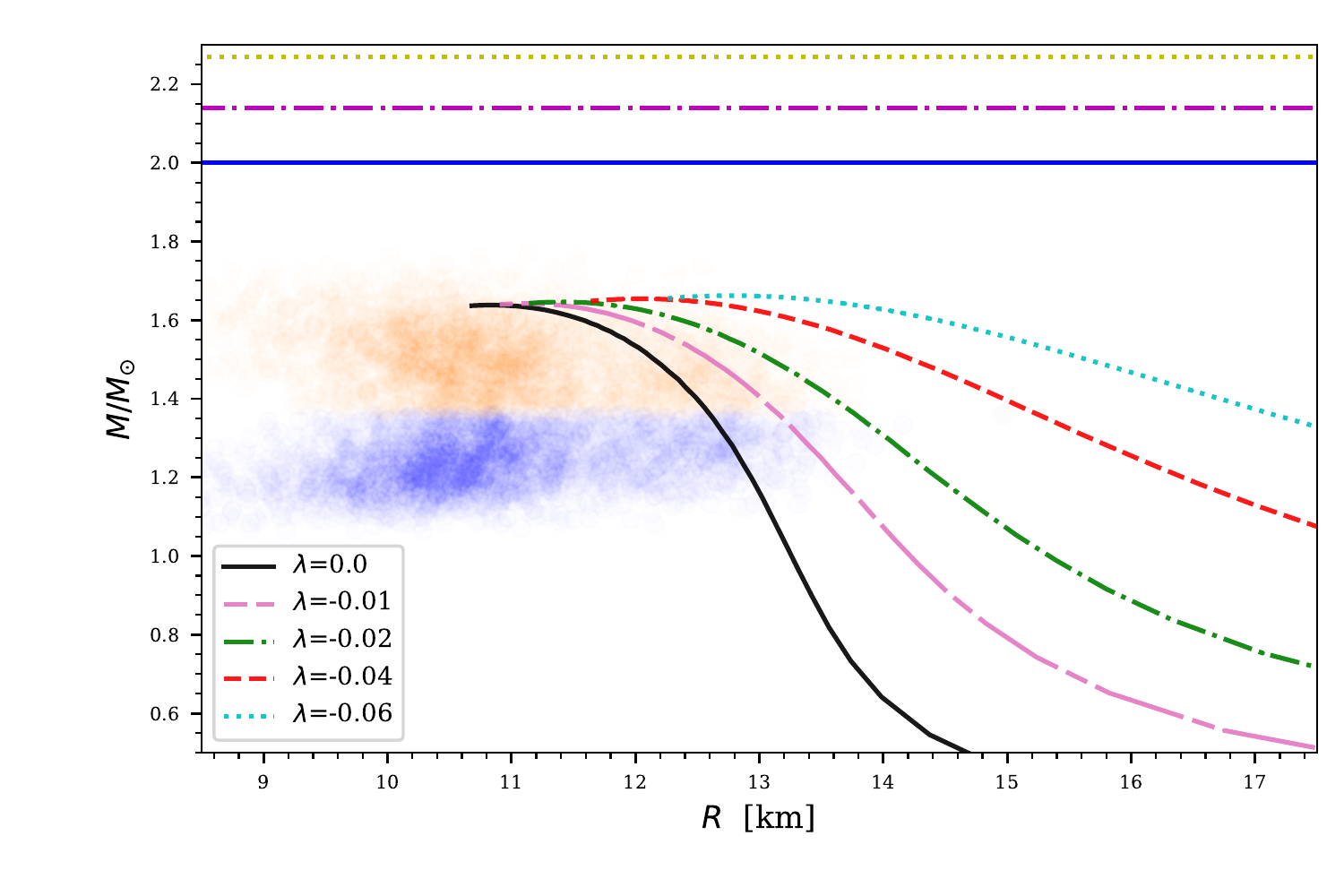}
      \caption{Z271s4 equation of state.}\label{fig:mr_z271s4}
    \end{subfigure}
    \caption{Mass-radius relation: On the upper left side the mass-radius relation for the BKA20 equation
      of state. On the upper right side the mass-radius relation for the BSR8 equation of state. On
      the lower left side the mass-radius relation for the IU-FSU equation
      of state. On the lower right side the mass-radius relation for the Z271S4 equation of state. It was
    considered five values of $\lambda$ in the mass-radius for each EoS, going from $\lambda=-0.06$ to 0.0, for $\lambda=0$, the theory retrieves general relativity. The blue and orange cloud region is the constraints for
      mass-radius from the GW170817 event, which was a merger of two neutron stars with an
      observation in the electromagnetic and gravitational spectrum. The blue continuous line at
      2.0~$M{\odot}$, the magenta dot-dashed line at 2.14~$M{\odot}$ and the yellow dot line at
      2.27~$M{\odot}$ represent the most massive
      pulsars observed up to now. The pulsar with 2.14~$M{\odot}$ has a 95.4\% credibility level.}\label{fig:mr_rmf}
  \end{figure}

In Figure~\eqref{fig:mr_rmf}, we present the study of the $f(\mathcal{R,T})$ hydrostatic equilibrium for
the relativistic mean-field models. We have chosen four EsoS from RMF models, which are
representative parametrizations of the set BKA, BSR, FSU, and Z271. These EsoS are well suitable in the context of tidal deformability in the GW170817 event. In
Figure~\eqref{fig:mr_bka20}, we have the parametrization BKA20; in Figure~\eqref{fig:mr_bsr8}, we have the
parametrization BSR8; in~\eqref{fig:mr_iufsu} the parametrization IU-FSU;~and in~\eqref{fig:mr_z271s4},
the last one, the Z271S4. As we can see, these parametrizations almost do not reach two solar masses if we consider GR only, being the worst case for Z271S4, which is under 1.8~$M_{\odot}$. The BKA20 is out of the LIGO-VIRGO region delimited by the gravitational wave detection. If one considers the contributions coming from $f(\mathcal{R,T})$ gravity, mass-radius values get even worsen, i.e.,
there are no values in this model for the $\lambda$ parameter. For the BSR8, the GR case is within a less dense
cloud region, and there are no values for the $\lambda$ also. The IU-FSU is in a more dense region, and it is possible to have a minimum value for $\lambda$ around $-0.01$; however,
neither GR nor $f(\mathcal{R,T})$ can reach two solar masses. These EsoS in the case
of GR and $f(\mathcal{R,T})$ theories of gravity do not predict a maximum mass upper to
2~$M_{\odot}$. Besides, if we consider these RMF hadronic models in $f(\mathcal{R,T})$ theory of gravity, the NS radii are out of the mass-radius region of LIGO-VIRGO observation except for the IU-FSU where the magnitude of $\lambda$ needs to be very small, less than $0.01$.

\subsection*{The crust}
As can we observe from figure~\eqref{fig:mr_apr} to~\eqref{fig:mr_rmf}, there is no significant enhancement in the mass of the compact star if we consider
the $f(\mathcal{R,T})$ theory with realistic equations of state: the maximum
increment was less than $1\%$. These new results are different from our previous
one~\cite{moraes/2016}, where we considered an EoS described by a polytrope, with $\Gamma=5/3$ and an EoS (using the MIT bag model)
for strange stars. By increasing the absolute value of the $\lambda$ parameter, we saw a significant increment in the mass of those stars. In another work~\cite{carvalho/2017}, we considered the Chandrasekhar EoS~\cite{chandrasekhar/1931,
  chandrasekhar/1935} for a white dwarf, and we observed an increment of up to 5\% in the white
dwarf mass. The limit of 5\% was related to
the limit of $\lambda$, which must be around $-4\times10^{-4}$. For values below that, the mass tends
to a plateau, and the radius would increase indefinitely.

As we could observer for white dwarfs, the main contribution from the
$f(\mathcal{R,T})$ model was regarding the stars' radius, i.e., there is an increment in the radius
for a decrease in the central density; we found the same result in this work considering realistic EsoS.

The physical reason for the difference relies on the presence of a crust, or more fundamentally due to
the low sound speed. When we model the stellar
structure, the models should consider the stellar core
and a crust, i.e., as one moves from the core to the surface, the density diminishes, and for a very low density, the EoS changes. The crust is generally described
by two layers, an inner and an outer crust. Atomic structure combined with the
nuclear theory is needed for the inner surface layer~\cite{haensel/2007}, while the outer layer
requires atomic structure combined with plasma physics in high density/temperature regimes. For
hadronic and hybrid stars, in general, it is used the Baym-Pethick-Sutherland (BPS)~\cite{baym/1971a} in this low-density regime.

Here, we considered two cases to describe the NS;~the first one was a star composed of a core and an inner crust only; the inner layer should have 1--2 km and a maximum density of $\sim0.5\rho_0$,
where $\rho_0$ is the nuclear saturation density. In the inner core, we used the SLy EoS (this EoS, and
in few cases the FPS one, are generally used do describe the inner crust, for details see the
Ref.~\cite{haensel/2007}). In the second case, we considered a core, an inner,
and an outer crust. A polytropic form describes the inner crust as
$p=A+B\rho^{4/3}$, where $A$ and $B$ are constants determined in the matching with the core and with the
outer layer. The BPS EoS describes the outer layer; for details, see the Ref.~\cite{lourenco/2020}.

\begin{figure}[ht]
    \centering
    \begin{subfigure}[b]{0.495\textwidth}
    \includegraphics[scale=0.47]{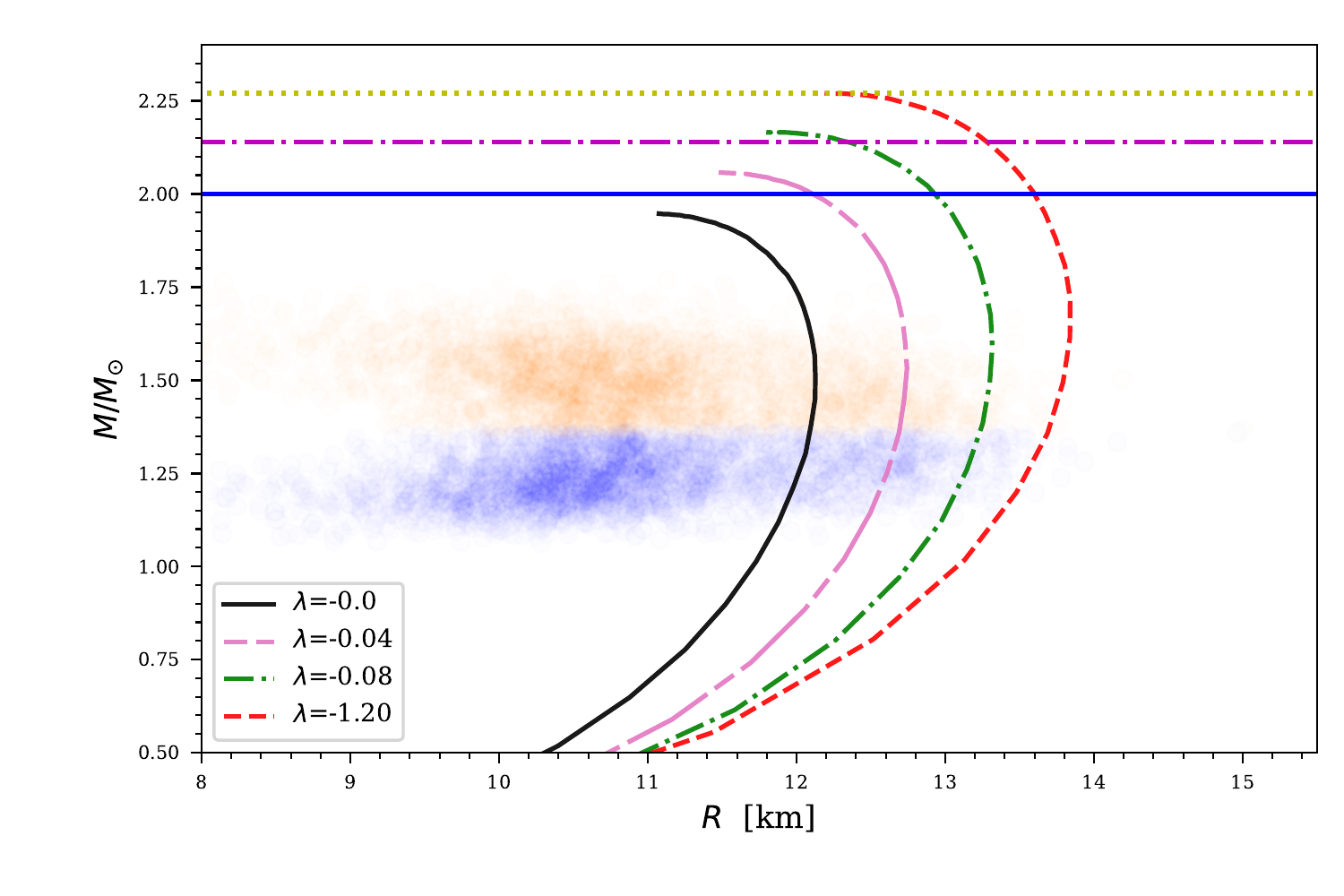}
      \caption{BKA20 equation of state.}\label{fig:mr_bka20_sc}
    \end{subfigure}
    \begin{subfigure}[b]{0.495\textwidth}
    \includegraphics[scale=0.47]{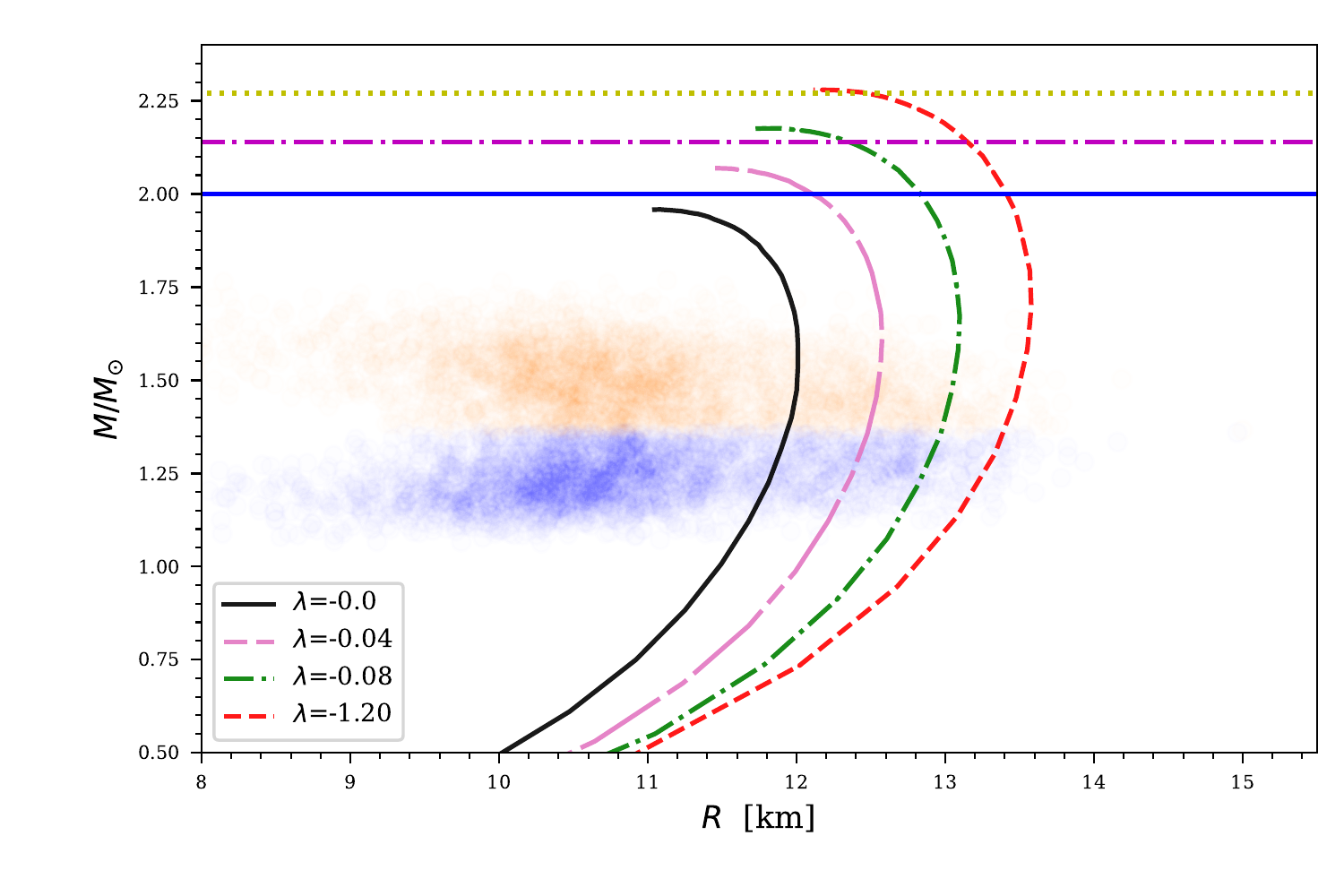}
      \caption{BSR8 equation of state.}\label{fig:mr_bsr8_sc}
    \end{subfigure}
    \caption{Mass-radius relation for bare stars: On the left side the mass-radius relation for the BKA20 equation
      of state. On the right side the mass-radius relation for the BSR8 equation of state. It was
    considered four values of $\lambda$ in the mass-radius for each EoS, going from $\lambda=-1.2$
    to 0.0, for $\lambda=0$, the theory retrieves general relativity. Here we considered the stellar
    structure only composed of a core, i.e., the star do not have a crust. The blue and orange cloud region is the constraints for
      mass-radius from the GW170817 event, which was a merger of two neutron stars with an
      observation in the electromagnetic and gravitational spectrum. The blue continuous line at
      2.0~$M{\odot}$, the magenta dot-dashed line at 2.14~$M{\odot}$ and the yellow dot line at
      2.27~$M{\odot}$ represent the most massive
      pulsars observed up to now. The pulsar with 2.14~$M{\odot}$ has a 95.4\% credibility level.}\label{fig:mr_sc}
  \end{figure}

If we consider only the core (Figure~\eqref{fig:mr_sc}), i.e., a bare star, one can have an increment in
the mass, as in the case of the NS and quark star, described by a simple polytropic and an MIT bag
model respectively~\cite{moraes/2016}, where the speed of sound is not reduced drastically near the surface of the
star. As we can see in Fig.~\eqref{fig:mr_sc}, one can have large absolute values of $\lambda$ compared to the one where we have to consider a crust.

In fact, in the results just presented, we found that using several relativistic and
non-relativistic models the small increment on the neutron star mass and the variation in the
stellar radius are almost the same for all the EsoS, which manifests that our results are very
insensitive to the EoS high density part of the star core in $f(\mathcal{R,T})$ gravity, in some
sense an unexpected result. It confirms that stellar structure changes in this alternative gravity theory depend almost only on the star crust, where the EoS is essentially the same for all the models.

\section{Discussion and Conclusion}\label{sec:dis}

In this paper, we obtained the mass-radius relationship within the $f(\mathcal{R,T})$ gravity for different
sets of EsoS with different parametrizations. It is the first time that the hydrostatic equilibrium equations are solved using realistic EsoS and having taken into account a joint constraint from massive pulsar and the gravitational wave event GW170817 (LIGO-VIRGO event) in the scope of $f(\mathcal{R,T})$ for the case $f(\mathcal{R,T})=R+2\lambda\mathcal{T}$. We took the stellar mass of $2~M_{\odot}$ as a benchmark and the radius from the LIGO observation as a threshold for the allowed EsoS and the values of the lambda parameter in the hydrostatic equilibrium equations. The EsoS used are from a wide range of softest/stiffness, which can be constrained by gravitational and electromagnetic events simultaneously. Some of them are based on experimental nuclear physics, making use of many-body
computations and other kinds of state-of-the-art calculations.

Our work shows that the main contribution from this $f(\mathcal{R,T})$ model is an increase in the stars' radius, i.e., we can have grander stars with smaller central density, as we have already shown in our
previous works~\cite{moraes/2016, carvalho/2017} for neutron stars (considering simplest EsoS) and
white dwarfs. In these previous works, for the case
of the NS (a polytrope with $(\Gamma=5/3)$) and quark star (an MIT bag model with $a=0.28$ and
$B=60~\mathrm{MeV/fm^3}$), it was found that the maximum mass increases with the increment of the absolute
value of
$\lambda$, the same behaviour we observed in WDs using the Chandrasekhar
EoS~\cite{chandrasekhar/1931, chandrasekhar/1935}, where we found a very slight increase, less than 5\%, in the mass, for the lowest value
allowed, $\lambda=-4\times10^{-4}$. However, with our new results using realistic EsoS, we show that the increment in the mass is less than 1\%.

Looking for the physical reason for the difference between these new results and the previous ones, we found that the neutron star crust
is responsible for this discrepancy. When we considered the crust, the maximum mass
increase is minimal; this is due the term $[1+\lambda/(8\pi+2\lambda)(1-d\rho/dp)]$ in~\eqref{tov2},
where
$(1-d\rho/dp) = (1-1/v^{2}_s)$ considering the EoS sound speed definition\footnote{$d\rho/dp=1/v^{2}_s(p)\geq 1/c^{2}$}. In the
causal limit, theory's contribution is zero, while $v^{2}_{s}\rightarrow 0$ the
term goes to $-\infty$. With the crust, the speed of sound is drastically reduced in the
outer layers and it is not possible to have a significant enhancement in the mass. This will be valid
for any equilibrium equation involving the sound speed, likewise for the works described
in \S2.3.11 of the Ref.~\cite{olmo/2020} where the majority have considered bare stars.

The speed of sound determines the parameter $\lambda$ in~\eqref{tov2} i.e., the softest/stiffness of
the EsoS, so the allowed values are $-4\pi<\lambda\leq0$ for $0<v_s\leq1$; whereas for $0<\lambda$
or $\lambda<-4\pi$ the condition $\lambda/(3\lambda+8\pi)<v_s$ must be satisfied. In the case of a MIT bag
model, where the sound speed is constant and satisfies the last condition, it is possible to have positive values of $\lambda$.

 Thus, because of the NS stellar structure considered here, which has a crust with a very soft
   EoS and, it is not possible to have big absolute values for the $\lambda$ parameter. Since this
   parameter is responsible for the strength of the new contributions coming from the theory and the corresponding masses and radii of NSs in $f(\mathcal{R,T})$, we can conclude that this alternative gravity cannot improve GR results in order to raise the two solar mass threshold.

Our results show that only the following EsoS are suitable within the
$f(\mathcal{R,T})= \mathcal{R}+\lambda\mathcal{T}$ model: APR3--4, WFF1--2, ENG, and MPA1. From these EsoS
and the joint constrains from the massive pulsar and the GW170817 event, we could deduce that
the minimum allowed value for the $\lambda$ parameter would be around $-0.02$ for neutron stars, and
this would increase the maximum mass less than 1\% for these stars.
Moreover, since
  $\lambda$ needs to be so small, not only the star mass is almost unchanged but also the star radius cannot become very large as in previous studies, and
  its increase compared to general relativity results is limited to be around 2.5--3.6\% in all the
  cases considered. The conclusion concerning the crust NS effect implying very small values of
  $\lambda$ does not depend on the form we have used to the $f(\mathcal{T})$, since for any other
  function of the trace of the energy-momentum tensor we could have chosen we would always have the inverse sound speed dependence $d\rho/dp$
  in the relativistic hydrostatic equilibrium for the neutron star, see Eq. (3) in Ref.~\cite{carvalho/2020b}.

  Let us stress that with the purpose of checking the possibility of attaining higher
    maximum masses for NSs, the present analysis may be extended to incorporate $R^2$ correction
    terms in gravitation, what was done, for instance, in \cite{mathew/2020, astashenok/2017,
      resco/2016}. Note that this incorporation could be a possibility to alleviate the $f(R)$ gravity shortcomings mentioned in the Introduction.

Finally, we would like to emphasize that our results indicate that conclusions obtained from compact stars studies done in alternative
  theories of gravity without using realistic EsoS to describe correctly the neutron star interior can be unreliable.

\acknowledgments
RL has been supported by U.S. Department of Energy (DOE) under grant DE-FG02-08ER41533, the LANL Collaborative
Research Program by Texas A\&M System National Laboratory Office and Los Alamos National Laboratory, CAPES/PDSE/88881.134089/2016--01 and Conselho Nacional de
Desenvolvimento Cient\'\i fico e Tecnol\'ogico (CNPq) process 141157/2015--1. PHRSM would like to thank CAPES for financial support. MGBA acknowledges CNPq Project 150999/2018--6 for financial support. The authors acknowledge the FAPESP Thematic Project 2013/26258--4. This work is also a part of the project INCT-FNA Proc. No. 464898/2014--5. OL and MD thank the support from CNPq, under Grants No. 310242/2017--7 (OL), No. 406958/2018--1 (OL), and No. 433369/2018--3 (MD), and from FAPESP under the thematic Project No. 2017/05660--0.
WP thank the support from CNPq, under grants No. 438562/2018--6 and No. 313236/2018--6 and CAPES
under the grant 88881.309870/2018--01.  M.M. acknowledges also Capes and CNPq for the financial
support. We thank both the Referee and Scientific Editor for valuable help in optimizing the
presentation of our paper as well as for the detailed and constructive discussions.

\bibliography{lib}
\bibliographystyle{JHEP}
\end{document}